\documentclass[lettersize,journal]{IEEEtran}

\usepackage{comment}
\usepackage{amssymb}

\usepackage[utf8]{inputenc}
\usepackage{xcolor}
\usepackage{enumitem}
\usepackage[hypertexnames=false,bookmarks=false]{hyperref}
\usepackage{color}
\usepackage{listings}
\usepackage[T1]{fontenc}

\usepackage{multirow}
\usepackage{url}
\usepackage{colortbl}
\usepackage{amsmath}
\usepackage{float}
\usepackage{graphicx}
\usepackage{algorithmic}
\usepackage{listings}
\usepackage{amsfonts}
\usepackage[linesnumbered, ruled, vlined]{algorithm2e}
\usepackage{longtable}
\usepackage{acronym}
\usepackage{tabularx}
\usepackage{booktabs}
\usepackage{tikz}
\usepackage{graphicx}
\usepackage{subcaption}
\usetikzlibrary{arrows.meta,positioning,shapes.geometric}
\usepackage{url}
\usepackage{todonotes}

\definecolor{lightgray}{rgb}{.9,.9,.9}
\definecolor{darkgray}{rgb}{.4,.4,.4}
\definecolor{darkgreen}{rgb}{0, 0.39, 0.00}
\definecolor{Gray}{gray}{0.7}
\definecolor{codegreen}{rgb}{0,0.6,0}
\definecolor{codegray}{rgb}{0.5,0.5,0.5}
\definecolor{codepurple}{rgb}{0.58,0,0.82}
\definecolor{backcolour}{rgb}{0.95,0.95,0.92}

\lstdefinestyle{mystyle}{
    backgroundcolor=\color{backcolour},   
    commentstyle=\color{codegreen},
    keywordstyle=\color{magenta},
    numberstyle=\tiny\color{codegray},
    stringstyle=\color{codepurple},
    basicstyle=\ttfamily\footnotesize,
    breakatwhitespace=false,         
    breaklines=true,                 
    captionpos=b,                    
    keepspaces=true,                 
    numbers=left,                    
    numbersep=5pt,                  
    showspaces=false,                
    showstringspaces=false,
    showtabs=false,                  
    tabsize=2
}

\acrodef{AEAD}{Authenticated Encryption with Associated Data}
\acrodef{AES}{Advanced Encryption Standard}
\acrodef{AHRS}{Attitude and Heading Reference System}
\acrodef{API}{Application Programming Interface}
\acrodef{CCM}{Counter with CBC-MAC}
\acrodef{CIAAN}{Confidentiality, Integrity, Availability, Authenticity, Non-Repudiation}
\acrodef{CRC}{Cyclic Redundancy Check}
\acrodef{CW}{Clockwise}
\acrodef{CCW}{Counter-Clockwise}
\acrodef{DDoS}{Distributed Denial of Service}
\acrodef{DoS}{Denial of Service}
\acrodef{DShot}{Digital Shot}
\acrodef{DUDE-IDS}{Distributed UAV Delay-based Intrusion Detection System}
\acrodef{EKF}{Extended Kalman Filter}
\acrodef{EKF3}{Enhanced Kalman Filter Version 3}
\acrodef{ESC}{Electronic Speed Controller}
\acrodef{GCM}{Galois/Counter Mode}
\acrodef{GCS}{Ground Control Station}
\acrodef{GPS}{Global Positioning System}
\acrodef{HAL}{Hardware Abstraction Layer}
\acrodef{HUD}{Heads-Up Display}
\acrodef{I2C}{Inter-Integrated Circuit}
\acrodef{IDS}{Intrusion Detection System}
\acrodef{IMU}{Inertial Measurement Unit}
\acrodef{LSTM}{Long Short-Term Memory}
\acrodef{MAVLink}{Micro Air Vehicle Link}
\acrodef{MEMS}{Micro-Electro-Mechanical Systems}
\acrodef{MUVIDS}{MAVLink-based UAV Intrusion Detection System}
\acrodef{NED}{North-East-Down}
\acrodef{OCB}{Offset Codebook Mode}
\acrodef{PID}{Proportional-Integral-Derivative}
\acrodef{PWM}{Pulse Width Modulation}
\acrodef{RC}{Remote Control}
\acrodef{RTL}{Return to Launch}
\acrodef{SDA}{Sensor Deprivation Attack}
\acrodef{SITL}{Software-in-the-Loop}
\acrodef{SIV}{Synthetic Initialization Vector}
\acrodef{SPI}{Serial Peripheral Interface}
\acrodef{TCP}{Transmission Control Protocol}
\acrodef{UAV}{Unmanned Aerial Vehicle}
\acrodef{USB}{Universal Serial Bus}

\begin{document}


\title{Control-Aware Manipulation of ArduPilot via Legitimate MAVLink Commands: Simulation and Hardware Validation}

 \author{Feras Benchellal, Lotfi Ben Othmane, Yasaswini Konapalli, Cihan Tunc, Bharat Bhargava

\thanks{F. Benchellal, L. Ben Othmane, Y. Konapalli, and C. Tunc are with the University of North Texas, Denton, TX, USA.}
\thanks{B. Bhargava is with Purdue University, West Lafayette, IN, USA.}
\thanks{Manuscript received [DATE]; revised [DATE].}}




\maketitle

\begin{abstract}
This paper investigates control-aware attacks against ArduPilot-based \acp{UAV}, in which an adversary exploits the sensitivity of flight-controller dynamics to parameter changes to cause loss of control and crashes. It describes six attacks that exploit interactions among multi-layer controllers by modifying \ac{PID} gains, altering \ac{EKF} estimation configuration, and violating failsafe assumptions, thereby forcing ArduPilot into unsafe operating conditions. We evaluate the attacks in \ac{SITL} simulation and validate them on a Pixhawk 2.4.8 hardware platform. The results show that short sequences of well-formed MAVLink messages can exploit controller sensitivity to parameter values and updates frequency, affecting controller states and degrading attitude stability, angular-rate behavior, trajectory tracking, and estimator health. We demonstrate that when multiple effects are combined, the vehicle can enter an unsafe state and crashes. These findings show that security gaps in input-parameter handling, command trust, and controller-state validation can be exploited to cause loss of control and crashes in \acp{UAV}.
\end{abstract}



\section{Introduction}
\label{sec:intro}
\acfp{UAV}, especially quadcopter drones, have been widely adopted for commercial, consumer, and military applications, including agriculture, infrastructure inspection, package delivery, emergency response, surveillance, and reconnaissance. This adoption has contributed to rapid growth in the global drone market, which is projected to increase from USD 83 billion in 2025 to USD 182 billion by 2033~\cite{GVR2026}. A key driver of this growth is that drones can operate autonomously or semi-autonomously in shared airspace at relatively low cost and without requiring complex supporting infrastructure. Nevertheless, drone security is a growing concern as physical or cyber attacks against drones can cause operational disruption or failure, financial loss, and even human safety. 

Research on \ac{UAV} security mainly focuses on four directions. First, protocol-level studies have shown that injected or replayed MAVLink packets can disrupt missions~\cite{marty2014vulnerability,kwon2018empirical,hamza2024mavlink}. Second, physical sensor attacks have demonstrated that acoustic or electromagnetic interference can corrupt \ac{MEMS} gyroscope and accelerometer measurements~\cite{son2015rocking,tu2018injected,trippel2017walnut}. Third, \acp{IDS} have been proposed to identify anomalies in communication patterns, software behavior, or flight dynamics~\cite{jeong2021muvids,tufekci2024intrusion,11140483,tan2025radd}. Fourth, cryptographic mechanisms have been proposed to secure the MAVLink channel~\cite{khan2022secure,tufekci2024enhancing}. 
These efforts improve \ac{UAV} security stack, but do not address the control-system consequences of accepted, well-formed commands.

For the control and management of drones, open-source ArduPilot autopilot software is generally preferred that runs on a flight-controller hardware to provide stabilization, navigation, mission execution, and failsafe management for \acp{UAV}~\cite{Ardupilot.org}. These systems also depend on the \acf{MAVLink} protocol for command-and-control communication because it is lightweight and efficient over low-bandwidth telemetry links~\cite{marty2014vulnerability,kwon2018empirical,hamza2024mavlink}. 
ArduPilot includes several security mechanisms~\cite{SecArdupilot2026} though their deployment could be constraints with overhead~\cite{hamza2024mavlink}. These mechanisms primarily follow a boundary-protection model: the \ac{GCS} is assumed to be trusted, the communication channel is protected through signing, and the firmware is protected from tampering through signature verification. In many deployments, however, commands may still be transmitted over the network in plaintext, authentication may be disabled or misconfigured, and the \acf{CRC} protects only against accidental transmission errors rather than malicious command injection~\cite{marty2014vulnerability,kwon2018empirical}.

ArduPilot uses a cascaded flight-control architecture in which position targets feed velocity controllers, velocity targets feed attitude controllers, attitude targets feed rate controllers, and rate-controller outputs ultimately drive the motors~\cite{KOTFM20206,ardupilot_copter_params}. It also uses \acf{EKF}, which fuses sensor measurements to estimate the vehicle state~\cite{10.1115}. Each layer in th control-cascade relies on commands, state estimates, and parameters received from other components. ArduPilot exposes many of these parameters through standard \ac{MAVLink} messages allowing operators to change them during flight. Thus, if an attacker modifies a \acf{PID} gain~\cite{minorsky1922directional} or an \ac{EKF} parameter through a well-formed \texttt{PARAM\_SET} message, ArduPilot may apply the new value at runtime unless parameter-locking protections are enforced. The autopilot does not evaluate whether the resulting closed-loop behavior remains safe under the current flight conditions. Consequently, corrupted control parameters can propagate through the cascade control loops and produce unsafe behavior, especially when updates are issued at carefully chosen times or frequencies.

\begin{figure}[htbp]
\centering
\includegraphics[width=.7\linewidth]{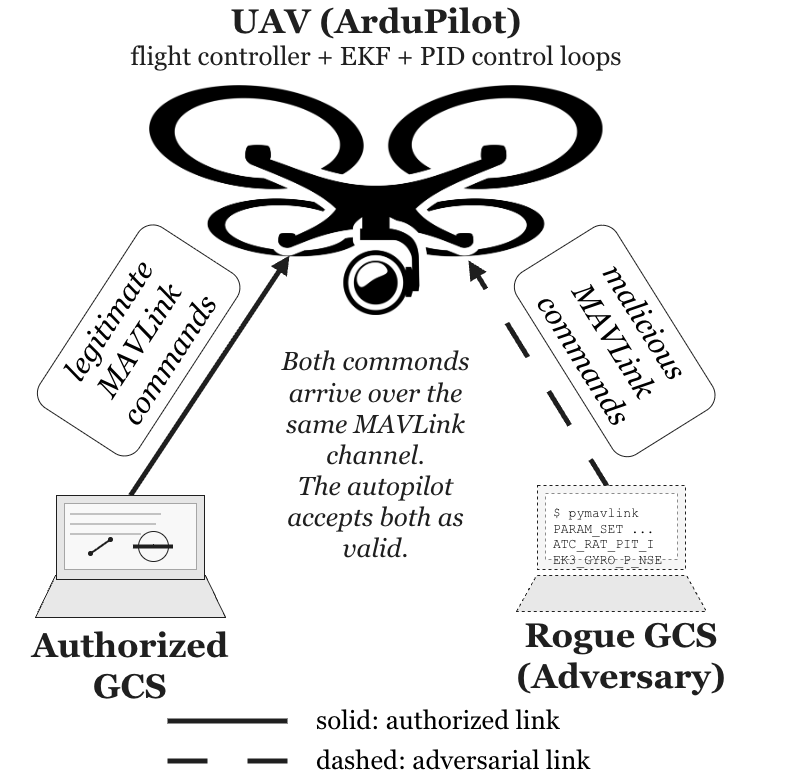}
\caption{Control-aware manipulation against a UAV.}
\label{fig:threat_overview}
\end{figure}

Prior works on \acp{IDS} for MAVLink-enabled and ArduPilot-based drones~\cite{11140483,jeong2021muvids,tufekci2024intrusion,tan2025radd} primarily focus on communication-level anomalies, software runtime behavior, or general flight anomalies and do not explicitly reason about command sequences manipulating Ardupilot's closed-loop control system. As a result, an adversary capable of placing well-formed packets on the communication channel, or controlling an authorized endpoint, may be able to issue commands that the autopilot accepts and executes without validating the paramters input. Figure~\ref{fig:threat_overview} illustrates this scenario: a legitimate \ac{GCS} and an adversary both send standard \ac{MAVLink} messages to the \ac{UAV} over telemetry channels. The parameter change commands sequence from the attacker can destabilize the drone, degrade tracking, bypass failsafe assumptions, or cause loss of control.

This paper investigates a class of attacks that we call \textit{control-aware attacks}. In these attacks, an adversary exploits the dynamics of ArduPilot’s control system using legitimate \ac{MAVLink} commands with maliciously chosen values exploiting the logic and dynamics of the control system. \ac{PID} controllers and estimation models such as the \ac{EKF} are inherently sensitive to their configuration parameters; changing these parameters can significantly affect stability, responsiveness, and state-estimation quality. Traditional network-boundary defenses do not address control-aware attacks targeting control loops, including the \ac{PID} controllers, the \ac{EKF} state estimator, and the cascaded architecture that translates high-level position commands into motor outputs. By exploiting the implicit trust among these components, we craft command sequences that push ArduPilot into dangerous states, including severe instability, loss of control, and crash. Existing protocol- and traffic-centered defenses can identify malformed packets, unauthorized traffic, or suspicious communication patterns; however, they are not well suited to preventing command sequences from inducing undesirable outcomes through malicious manipulation of the behaviors managed by the ArduPilot control system, as shown by this paper.

We consider two access models. In the \textit{channel-injection model}, MAVLink signing is absent or misconfigured, allowing the adversary to inject packets over the telemetry link. In the \textit{authorized-client model}, MAVLink authentication may be enabled, but the adversary controls a trusted endpoint, such as a compromised \ac{GCS} or companion computer. These capabilities could also arise from weak \ac{GCS} authentication or malware running on an authenticated ground-control system. Our attacks require only the ability to issue standard command and parameter messages. They do not require firmware modification, sensor spoofing, radio jamming, or packet flooding. 


We define two attack outcomes. \textit{Partial} success means measurable flight degradation, such as attitude oscillation, tracking error, angular-rate instability, or \ac{EKF} drift, without a crash.\footnote{These attacks may be interpreted as analysis of the sensitivity of the individual controllers as they do not cause malicious outcome.} \textit{Full} success means the vehicle reaches an unrecoverable state, triggers crash detection, have the operator lose control, or impacts the ground. We exclude \acf{DoS} attacks that rely on flooding the communication link, as those attacks are already well studied~\cite{kwon2018empirical}. Instead, we focus on targeted command sequences whose danger comes from control-system effects rather than network-volume effects.

This paper makes the following contributions:
\begin{enumerate}

\item We design six control-aware attacks against ArduPilot, each targeting a different subsystem or combination of subsystems: oscillation injection against the position controller, \ac{PID} gain manipulation against the pitch-rate controller, \ac{EKF} confusion through noise-parameter corruption, failsafe bypass through combined \ac{PID} corruption and position oscillation, yaw-axis destabilization through yaw-rate controller manipulation, and a combined multi-layer attack on the \ac{EKF}, rate controllers, and attitude controllers.
\item We validate all six attacks in both \ac{SITL} simulation and physical hardware demonstrations on a Pixhawk 2.4.8 platform. We quantify their effects using DataFlash telemetry logs that capture attitude stability, angular-rate variability, \ac{EKF} health, tracking error, and \ac{GPS} trajectory deviation.
\item We identify specific security gaps in ArduPilot's design. In the tested configuration, ArduPilot accepted selected runtime parameter updates without enforcing control-aware safety constraints on their closed-loop effects. These gaps include limited validation of parameter modifications, implicit trust in command sequences regardless of timing or magnitude, and the inability of existing failsafe mechanisms to detect control-aware manipulation.

\end{enumerate}

The novelty of this paper is not merely that unsafe parameter values can exist; rather, it lies in systematically mapping ArduPilot’s cascaded control architecture to attack primitives, constructing MAVLink protocol-compliant command sequences that can cause loss of operator control and vehicle crashes, and validating their control effects in both \ac{SITL} simulation and Pixhawk hardware.

The rest of the paper is organized as follows. Section~\ref{sec:controllers} describes ArduPilot's cascaded controller architecture and the mathematical models our attacks target. Section~\ref{sec:related} surveys related work. Section~\ref{sec:simulation} presents the six attacks and their SITL evaluation. Section~\ref{sec:validation} reports the hardware validation results on a Pixhawk 2.4.8. Section~\ref{sec:conclusion} concludes.

\section{ArduPilot Control Architecture}
\label{sec:control_architecture}
\label{sec:controllers}

\begin{figure}[!tb]
    \centering
    \includegraphics[width=1.0\linewidth]{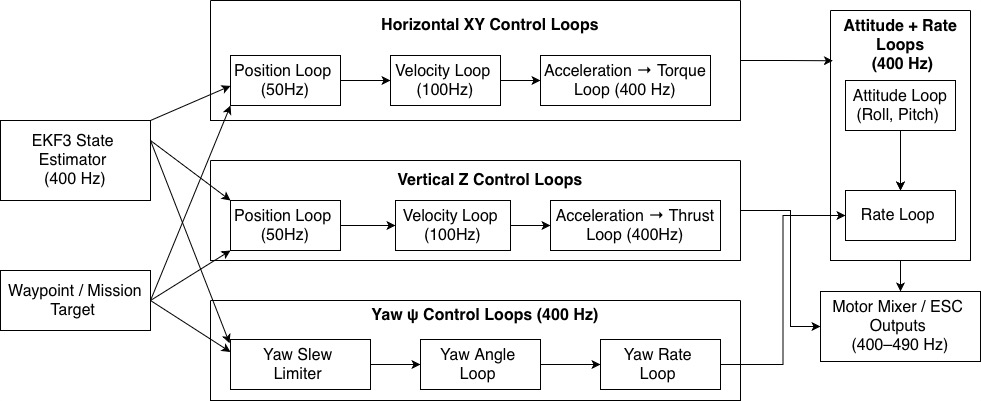}
    \caption{Reverse-engineered ArduCopter Control System~\cite{KOTFM20206}.}
    \label{fig:arducopter-flow}
\end{figure}

ArduPilot implements a layered control architecture that transforms high-level mission objectives into low-level motor commands through a cascade of nested control loops. This section summarizes the ArduPilot control architecture that we obtained by reverse-engineering the open-source ArduCopter code~\cite{Ardupilot.org} (ArduCopter is one of the specific firmwares designed for multirotors and helicopters within ArduPilot ecosystem) and reported in~\cite{KOTFM20206}. 
The reverse-engineered ArduCopter flow is demonstrated in Figure~\ref{fig:arducopter-flow}, which is the foundation of this work, and explained in this section. 

\subsection{System Overview and Reference Frames}

ArduCopter, as designed for multirotors (i.e., drones), adopts the \ac{NED} reference frame as its primary coordinate system, where the $x$-axis points the North, the $y$-axis points the East, and the $z$-axis points towards the Earth's center. This convention is widely adopted in aerospace applications and simplifies certain mathematical operations. All position, velocity, and acceleration vectors are expressed in this frame unless noted otherwise.

The vehicle state at time $t$ is characterized by position $p(t)$, velocity $v(t)$, and acceleration $a(t)$, each as three-dimensional vectors as shown in Eq.~\ref{eq:state_vectors} that flow through the control hierarchy from coarse position targets down to motor actuation commands.

\begin{equation}
p(t) = \begin{bmatrix} x \\ y \\ z \end{bmatrix} \text{ (cm)}, \quad
v(t) = \begin{bmatrix} v_x \\ v_y \\ v_z \end{bmatrix} \text{ (cm/s)}, \quad
a(t) = \begin{bmatrix} a_x \\ a_y \\ a_z \end{bmatrix} \text{ (cm/s}^2\text{)}
\label{eq:state_vectors}
\end{equation}

\subsection{Hierarchical Control Structure}

The control architecture implements a five-layer cascade that progressively transforms waypoints into motor-level \ac{PWM} signals. Each layer operates at a distinct frequency and produces setpoints for the layer below it, as shown in Figure~\ref{fig:arducopter-flow}.

\noindent\textbf{Layer 1: Position Control Loop (50 Hz).} The outermost loop processes position targets received via \ac{MAVLink} commands. It compares target positions against \ac{EKF} estimates and computes desired velocity setpoints using a \ac{PID} controller. This layer executes at 50 Hz. 

\noindent\textbf{Layer 2: Velocity Control Loop (100 Hz).} This layer takes velocity targets from the position controller and generates acceleration commands. It accounts for both tracking error and feed-forward terms.

\noindent\textbf{Layer 3: Attitude Control Loop (400 Hz).} Acceleration commands are transformed into attitude targets, specifically desired roll ($\phi$) and pitch ($\theta$) angles that produce the required thrust vector direction.

\noindent\textbf{Layer 4: Rate Control Loop (400 Hz).} The innermost feedback loop regulates angular velocities measured by gyroscopes, producing body-frame torque commands $\tau(t)$.

\noindent\textbf{Layer 5: Motor Mixing \& ESC Outputs Loop (400 - 490 Hz).} The final stage converts collective thrust and torques into individual motor commands through a mixing matrix. The mixer outputs \ac{PWM} or \ac{DShot} commands to \acp{ESC}.

This layered structure allows the modular analysis and decision making for a stable flight for drones. Nevertheless, this structure has a security consequence as the manipulated commands or parameters at any layer propagate through all inner layers below it; i.e., a corrupted velocity target from the position controller cascades into incorrect acceleration, attitude, and motor commands. The ArduPilot software has no mechanism to detect that the cascade has been compromised from the outside. 

\subsection{Sensor Fusion and State Estimation}

ArduPilot uses multiple sensors, such as \ac{IMU} for acceleration and angular rate, magnetometer for heading, barometer for altitude, and \ac{GPS} for position and velocity, and employs the \ac{EKF3}~\cite{10.1115} to fuse their measurements for a stable flight operation. 
The \ac{EKF} operates in a two-step prediction-correction cycle. 
In the prediction step, inertial measurements propagate the state forward, using Eq.~\ref{eq:ekf_predict}. 

\begin{equation}
\begin{aligned}
x_{t|t-1} &= f(x_{t-1}, u_{t-1}, w_{t-1}) \\
P_{t|t-1} &= F_t P_{t-1} F_t^T + Q_t
\end{aligned}
\label{eq:ekf_predict}
\end{equation}

\noindent where $x$ is the state vector (position, velocity, orientation, sensor biases), $f(\cdot)$ is the nonlinear state transition function, $F_t$ is its Jacobian, $P$ is the state covariance, and $Q_t$ is the process noise covariance.

When measurements $z_t$ arrive from slower sensors (\ac{GPS}, barometer, magnetometer), the correction step adjusts the predicted state using the Eq.~\ref{eq:ekf_update}. 

\begin{equation}
\begin{aligned}
K_t &= P_{t|t-1} H_t^T (H_t P_{t|t-1} H_t^T + R_t)^{-1} \\
x_t &= x_{t|t-1} + K_t (z_t - h(x_{t|t-1}))
\end{aligned}
\label{eq:ekf_update}
\end{equation}

\noindent where $K_t$ is the Kalman gain, $H_t$ is the measurement model Jacobian, and $R_t$ is the measurement noise covariance.

The \ac{EKF} is especially important in this work because the noise covariance parameters $Q_t$ and $R_t$ are configurable through \ac{MAVLink}. If we increase the process noise, the filter distrusts its own predictions and becomes overly reactive to sensor readings. If we increase measurement noise, it ignores sensor corrections and drifts. Either way, manipulating these parameters can make the filter converge to incorrect states while its internal covariance bounds still look healthy, which means the autopilot can continue to treat an unsafe state estimate as valid.

\subsection{Mathematical Control Models}
\label{sec:control_models}

All controllers in the ArduPilot cascade use variations of the discrete-time \ac{PID} control law~\cite{minorsky1922directional}. Understanding the general form, and how it applies to specific subsystems, helps identify which parameters to target and what behavior to expect when those parameters are modified.
While ArduPilot uses cascaded \ac{PID} as the dominant paradigm in open-source autopilots, the broader control literature has developed more sophisticated alternatives. Optimal and adaptive control approaches, including the reinforcement-learning-based optimal control framework developed by Lewis et al.~\cite{lewis2012optimal,lewis2009reinforcement}, offer principled handling of model uncertainty and disturbances that fixed-gain \ac{PID} cannot match. The vulnerability class identified in this study (runtime gain corruption via \texttt{PARAM\_SET}) would manifest differently in such systems but is not eliminated by them; extending control-aware attack analysis to optimal and adaptive controllers is a natural direction for future work.

\subsubsection{General PID Control Law}

For any controlled variable, error $e(t)$ for time $t$ is defined as the difference between target and measurement (Eq.~\ref{eq:generic_error}). 

\begin{equation}
e(t) = \text{target}(t) - \text{measurement}(t)
\label{eq:generic_error}
\end{equation}
and, using error information, the controller output is defined as $f$ using the Eq.~\ref{eq:pid_general}.

\begin{equation}
f(t) = k_P e(t) + k_I \sum_{i=0}^{t} e_i \Delta t + k_D \frac{e_t - e_{t-1}}{\Delta t} + k_{FF} \cdot \text{target}(t)
\label{eq:pid_general}
\end{equation}

\noindent where $k_P$, $k_I$, $k_D$ are the proportional, integral, and derivative gains, $\Delta t$ is the time step, and $k_{FF}$ is the feed-forward gain. The proportional term responds to current error, the integral term eliminates steady-state offsets by accumulating past error, the derivative term damps oscillations by reacting to error rate of change, and the feed-forward term anticipates required output from the target setpoint directly.

Every gain value ($k_P$, $k_I$, $k_D$, $k_{FF}$) is a tunable parameter exposed via \ac{MAVLink} \texttt{PARAM\_SET}. This is an intentional decision because operators must tune gains for different vehicle configurations (e.g., quadcopter or hexacopter or even autonomous cars). However, it also means that an attacker with parameter-write access can modify these values at runtime and the autopilot software applies the new values immediately.

\subsubsection{Position and Velocity Control}

Horizontal motion is governed by two nested loops: The outer position loop computes velocity targets from position error, and the inner velocity loop computes acceleration targets from velocity error. Both follow the \ac{PID} structure in Eq.~\ref{eq:pid_general}. 
The position controller computes error using the Eq.~\ref{eq:pos_error}:
\begin{equation}
e_{p,xy}(t) = p_{d,xy}(t) - p_{c,xy}(t)
\label{eq:pos_error}
\end{equation}
\noindent where $p_{d,xy}$ is the desired position (from \ac{MAVLink} commands, adjusted for \ac{EKF} offsets) and $p_{c,xy}$ is the current estimated position. The \ac{PID} output becomes the velocity target for the inner loop. In ArduPilot, the relevant gains are \texttt{PSC\_POSXY\_P}, \texttt{PSC\_POSXY\_I}, and \texttt{PSC\_POSXY\_D}.

The velocity controller then takes this velocity target and produces an acceleration command, as shown in Eq.~\ref{eq:vel_to_accel}.
\begin{equation}
\begin{aligned}
a_{d,xy}(t) &= k_{P_v,xy} e_{v,xy}(t)
+ k_{I_v,xy} \sum_{i=0}^{t} e_{v,xy}(i) \Delta t \\
&\quad + k_{D_v,xy} \frac{e_{v,xy}(t) - e_{v,xy}(t-1)}{\Delta t}
\end{aligned}
\label{eq:vel_to_accel}
\end{equation}
with gains from \texttt{PSC\_VELXY\_P}, \texttt{PSC\_VELXY\_I}, and \texttt{PSC\_VELXY\_D}. The resulting acceleration is bounded by \texttt{PSC\_ACCZ\_MAX} to prevent commands that would exceed actuator limits.
Setting position gains too high causes rapid velocity commands that saturate the inner loop and induce oscillation. Setting integral gains too high causes windup where accumulated error drives the vehicle past its target and into oscillation around it. These dynamics are the basis for the attacks described in Section~\ref{sec:simulation}.

Altitude control follows the same cascaded structure along the \ac{NED} $z$-axis with an additional thrust conversion step. The vertical acceleration command must be converted to a normalized thrust value using Eq.~\ref{eq:thrust_compensation}. 

\begin{equation}
T_{out}(t) = \frac{T_{hover} + T_{in}(t)}{\cos \alpha(t)}
\label{eq:thrust_compensation}
\end{equation}
\noindent 
where $T_{hover}$ is the steady-state hover thrust (parameter \texttt{MOT\_THST\_HOVER}, typically 0.3--0.5), $T_{in}$ is the \ac{PID} output, and $\alpha(t)$ is the vehicle tilt angle. The cosine term compensates for the fact that tilting the vehicle reduces the vertical component of thrust. This coupling means aggressive horizontal maneuvers require more thrust to hold altitude, and if motors saturate, the vehicle loses both horizontal and vertical control authority simultaneously.

The $T_{hover}$ parameter can be set through \texttt{PARAM\_SET}. Setting it too low makes the vehicle think it needs less thrust than it actually does, causing it to descend. Setting it too high makes it climb. In either case, a single parameter change disrupts the relationship between commanded acceleration and actual thrust output.

\subsubsection{Attitude and Rate Control}

The attitude and rate controllers operate at 400 Hz, the fastest loops in the system, due to their importance in a stable flight operation. The attitude controller converts acceleration demands into angular targets, and the rate controller tracks those targets by commanding motor torques. 
To achieve horizontal accelerations $a_{t,x}$ and $a_{t,y}$, the vehicle must tilt its thrust vector. For small angles, the desired roll and pitch are calculated using Eq.~\ref{eq:accel_to_angle}, 

\begin{equation}
\phi(t) \approx \frac{a_{t,x}(t)}{g}, \quad \theta(t) \approx -\frac{a_{t,y}(t)}{g}
\label{eq:accel_to_angle}
\end{equation}
where $g \approx 980$ cm/s$^2$. ArduPilot uses full quaternion math for arbitrary attitudes, but this approximation shows the core relationship: The horizontal acceleration scales directly with tilt angle. 
ArduPilot represents orientation using quaternions $q = [q_w, q_x, q_y, q_z]$ to avoid gimbal lock. The attitude error between target $q_t$ and current $q_b$ is calculated using Eq.~\ref{eq:quat_error}.

\begin{equation}
e_{ang}(t) = \text{axisangle}(q_t(t) \otimes q_b^{-1}(t))
\label{eq:quat_error}
\end{equation}

A proportional controller maps this error to target angular rates using Eq.~\ref{eq:att_to_rate}.
\begin{equation}
\boldsymbol{\omega}_t(t) = k_{P,ang} \odot e_{ang}(t)
\label{eq:att_to_rate}
\end{equation}

\noindent where $k_{P,ang}$  comes from parameters of \texttt{ATC\_ANG\_RLL\_P}, \texttt{ATC\_ANG\_PIT\_P}, and \texttt{ATC\_ANG\_YAW\_P}.
The rate controller is the innermost feedback loop and produces the actual torque commands sent to motors (using Eq.~\ref{eq:rate_pid}).

\begin{equation}
\begin{aligned}
\boldsymbol{\tau}(t) &= k_{P,\omega} e_\omega(t)
+ k_{I,\omega} \sum_{i=0}^{t} e_\omega(i) \Delta t \\
&\quad + k_{D,\omega} \frac{e_\omega(t) - e_\omega(t-1)}{\Delta t}
+ k_{FF,\omega} \boldsymbol{\omega}_t(t)
\end{aligned}
\label{eq:rate_pid}
\end{equation}

\noindent where $e_\omega(t) = \boldsymbol{\omega}_t(t) - \boldsymbol{\omega}_c(t)$ is the rate error. The gains come from \texttt{ATC\_RAT\_RLL\_P/I/D/FF}, \texttt{ATC\_RAT\_PIT\_P/I/D/FF}, \texttt{ATC\_RAT\_YAW\_P/I/D/FF}.
These rate controller gains are among the most sensitive parameters in the entire system. Even small perturbations to well-tuned gains can move the system toward its stability boundary. Too little $k_{P,\omega}$  causes vehicle to responds sluggishly and cannot reject disturbances. Too much $k_{P,\omega}$ causes oscillation violently as the controller tries to balance sensor noise and structural vibrations. Our \ac{PID} gain manipulation attacks directly exploit this sensitivity.

Yaw control adds slew rate limiting to prevent abrupt heading changes with Eq.~\ref{eq:yaw_rate_limit}. 

\begin{equation}
\dot{\psi}_t(t) = \text{clip}(\dot{\psi}_{cmd}(t), -\dot{\psi}_{max}, \dot{\psi}_{max})
\label{eq:yaw_rate_limit}
\end{equation}
\noindent where $\dot{\psi}_{max}$ is set by \texttt{RATE\_Y\_MAX}. This filtering creates an interesting dynamic for our oscillation attacks: Rapidly alternating yaw commands get smoothed by the slew limiter, but the vehicle still accumulates heading drift over time because the filtering is not symmetric in both directions.

\subsubsection{Motor Mixing}

The mixing stage converts thrust $T_{out}$ and torques $\boldsymbol{\tau}$ into individual motor commands. For a quadcopter in X-configuration, Eq.~\ref{eq:motor_mix} demonstrates the normalized motor thrusts, as follows. 

\begin{equation}
\mathbf{u}_m(t) = \begin{bmatrix} 1 \\ 1 \\ 1 \\ 1 \end{bmatrix} T_{out}(t) + M \boldsymbol{\tau}(t)
\label{eq:motor_mix}
\end{equation}

\noindent where $\mathbf{u}_m = [u_1, u_2, u_3, u_4]$ are normalized motor thrusts in $[0,1]$ and $M$ is the geometry-dependent mixing matrix. If any motor command exceeds $[0,1]$ bounds, ArduPilot applies proportional desaturation; i.e., it scales all commands values down while preserving their relative differences. This keeps the vehicle controllable at the cost of reduced total thrust.

Motor saturation is where many of our attacks ultimately manifest their effects. When corrupted parameters cause controllers to demand more torque or thrust than the motors can deliver, the mixer saturates. Sustained saturation means the vehicle cannot track its targets, errors accumulate through the cascade, integral terms wind up, and the error compounds until a failsafe triggers or the vehicle enters an unrecoverable state.

\subsection{Control Loop Dynamics}
\label{sec:loop_dynamics}

\subsubsection{Multi-Rate Cascade and Timing}

The cascade operates across four time scales, as mentioned previously: position control at 50 Hz, velocity at 100 Hz, attitude and rate at 400 Hz, and motor output at 400 - 490 Hz. Inner loops run faster than outer loops by factors of 2 - 10$\times$ to maintain time-scale separation, which is standard practice in control theory. Each fast inner loop appears quasi-steady to the slower outer loop above it. 

This separation has a security implication: A manipulated parameter in a fast inner loop can destabilize that loop without the slower outer loops noticing immediately. The outer loops keep commanding setpoints that the inner loop can no longer track. The result is a delayed instability that shows up as unexplained drift or oscillation at the position level while the actual root cause is in the attitude or rate dynamics. By the time the outer loops react, the situation may already be unrecoverable. 
A position command takes roughly 20 - 30~ms to propagate through the entire cascade to the motors, purely from computational sequencing. This latency is important for our attacks because rapidly oscillating commands must be timed relative to this propagation delay to achieve maximum destabilizing effect. 

\subsubsection{Inter-Loop Coupling and Saturation}

Although the cascade appears hierarchical, significant coupling exists between axes. Three mechanisms are directly relevant to our attacks:

\noindent\textbf{Thrust-Attitude Coupling.} Eq.~\ref{eq:thrust_compensation} shows that thrust must increase as tilt angle grows to maintain the vertical component. During aggressive horizontal commands, thrust can exceed motor capacity. When this happens the vehicle loses both altitude hold and position tracking simultaneously, which is the tumbling behavior we observe in several of our attacks.

\noindent\textbf{Yaw-Roll Coupling.} Yaw control relies on differential motor speeds between clockwise and counter-clockwise rotors. During combined roll and yaw commands, the mixer cannot satisfy both torque demands if motors saturate. ArduPilot prioritizes roll/pitch over yaw, meaning our yaw oscillation attacks cause heading drift while roll/pitch stability is preserved, at least until the situation gets bad enough to overwhelm everything.

\noindent\textbf{Integral Windup Cascade.} Each \ac{PID} controller accumulates integral error. If an inner loop saturates, error persists in the outer loop, of which integral keeps growing. When saturation eventually releases, the wound-up integral drives a large overshoot. In a three-layer cascade, windup can compound across all layers simultaneously. ArduPilot has anti-windup logic, but our parameter manipulation attacks are able to overwhelm these protections.

\subsection{Failsafe and Safety Mechanisms}
\label{sec:failsafes}

ArduPilot has extensive failsafe mechanisms for \ac{GPS} loss, \ac{RC} loss, low battery, excessive \ac{EKF} variance, and crash detection with configurable thresholds and response actions (continue, \ac{RTL}, land, disarm), all adjustable through \texttt{PARAM\_SET}. 
However, such a configurability introduces security weaknesses. For example, raising thresholds can prevent fault detection, letting the vehicle fly with degraded state estimation. Lowering thresholds causes false triggers from normal transients and noise. Changing the response action itself turns a safety mechanism into an attack vector, for example, setting the \ac{GPS} failsafe action to ``Land'' causes the vehicle to descend immediately whenever it loses \ac{GPS} lock, even briefly.

Crash detection is particularly an interest of this study. ArduPilot declares a crash when several conditions occur simultaneously for more than 2 seconds: lean angle exceeding 15$^\circ$, acceleration below 3~m/s$^2$, thrust error above 30$^\circ$, and horizontal velocity below 10~m/s. Our combined attacks that push the vehicle into sustained high lean angles can trigger this logic, causing ArduPilot to disarm the motors mid-flight.

The \ac{EKF} has its own internal health checks based on innovation variance (the difference between predicted and actual measurements). The parameter \texttt{FS\_EKF\_THRESH} sets the threshold for declaring the \ac{EKF} unhealthy. Our \ac{EKF} confusion attacks manipulate noise parameters to push innovation variance close to this threshold, creating a situation where the state estimates are degraded but the failsafe has not triggered yet, a dangerous gray zone where ArduPilot is flying on bad information but does not know it.

\subsection{MAVLink Commands Used}
 
Our attacks rely on a specific subset of MAVLink messages. These are standard messages any ground control station uses during normal operations. The most important is \texttt{PARAM\_SET}, which writes a new value to any autopilot parameter. We use it to modify PID gains, EKF noise thresholds, failsafe limits, and other configuration values at runtime. \texttt{SET\_POSITION\_TARGET\_LOCAL\_NED} and \texttt{SET\_POSITION\_TARGET\_GLOBAL\_INT} send position and velocity targets. \texttt{SET\_ATTITUDE\_TARGET} directly specifies desired attitude. \texttt{COMMAND\_LONG} handles arming, takeoff, and mode changes. \texttt{HEARTBEAT} keeps the GCS connection alive. In the tested configuration, ArduPilot accepts these messages without control-aware contextual validation; that behavior is the property our attacks exploit.

\section{Related Work}
\label{sec:related}
 
\noindent{\bf MAVLink Protocol Vulnerabilities.} The MAVLink protocol has been a primary entry point for UAV security research. Marty~\cite{marty2014vulnerability} was among the first to assess MAVLink vulnerabilities, demonstrating on the ArduPilot Mega 2.5 that unencrypted, unauthenticated messages allow injection, hijacking, and denial-of-service against operational UAVs. Kwon et al.~\cite{kwon2018empirical} confirmed empirically that a remote adversary can disable missions or force UAVs to hover, requiring only basic knowledge of the MAVLink message format. Hamza et al.~\cite{hamza2024mavlink} surveyed the protocol's evolution and current countermeasures, noting that MAVLink 2.0's optional signature field is still rarely deployed because key distribution is operationally inconvenient.
 
\noindent{\bf Cryptographic Solutions.} Khan et al.~\cite{khan2022secure} proposed a packet-level encryption scheme combining a Caesar cipher with custom character mapping and a synchronized key list. Tufekci et al.~\cite{tufekci2024enhancing} proposed authenticated encryption with associated data (AEAD) schemes, including ChaCha20-Poly1305, AES-GCM-SIV, AES-OCB3, and AES-CCM, that encrypt the payload while authenticating the header and provide replay protection through sequence-number authentication. These approaches strengthen channel security but do not address the class of attacks studied here, where commands are well-formed and authorized.
 
\noindent{\bf Physical Sensor Attacks.} Beyond protocol vulnerabilities, researchers have shown that even cryptographically secure command channels do not protect against attacks on the physical sensors themselves. Son et al.~\cite{son2015rocking} demonstrated that \ac{MEMS} gyroscopes in commercial drones are vulnerable to acoustic resonance attacks, causing rotor commands to saturate and rendering the drone uncontrollable. Tu et al.~\cite{tu2018injected} extended this to fine-grained adversarial control over inertial sensor outputs. Trippel et al.~\cite{trippel2017walnut} achieved similar control over \ac{MEMS} accelerometers via signal-conditioning vulnerabilities.
 
\noindent{\bf Sensor Fusion Vulnerabilities.} Shen et al.~\cite{shen2020drift} performed the first security analysis of multi-sensor fusion-based localization in autonomous driving, exploiting transient periods of elevated state uncertainty in a Kalman-filter implementation. Tu et al.~\cite{tu2019flight} addressed the related challenge of recovering flight when redundant \acp{IMU} are simultaneously compromised by acoustic resonance.
 
\noindent{\bf Sensor Deprivation and Reconfiguration Attacks.} Erba et al.~\cite{erba2024sensor} introduced sensor deprivation attacks, which manipulate drone behavior by reconfiguring onboard sensors via unauthenticated \ac{I2C}/\ac{SPI} \acp{API} rather than continuously injecting false data. A single malicious reconfiguration message persistently alters sensor behavior even after the attacker leaves range. This work is conceptually closest to ours: both exploit reconfiguration rather than continuous spoofing. Our attacks operate one abstraction layer higher, on the autopilot's parameter store, which is reachable from any authenticated MAVLink client.
 
\noindent{\bf Intrusion Detection.} Jeong et al.~\cite{jeong2021muvids} proposed \ac{MUVIDS}, a network-level \ac{IDS} for MAVLink-enabled vehicles that detects flooding and false-injection patterns. Tufekci~\cite{tufekci2024intrusion} presented \ac{DUDE-IDS}, a multi-layered IDS combining flow-based supervised learning at the network layer with LSTM-based sequence learning at the control layer. These systems detect protocol anomalies and unusual message sequences, but they may miss attacks that use properly formatted commands at normal rates. In such cases, malicious behavior arises from the timing, magnitude, and combination of values rather than from a protocol violation.
 
\noindent{\bf Software Reliability of Open-Source Autopilots.} The ArduPilot project was subjected to monthly Coverity scans for several years, identifying hundreds of defects, mostly memory-related~\cite{ardupilot2019coverity}. Wang et al.~\cite{wang2021exploratory} conducted a large-scale empirical study of UAV autopilot bugs from PX4 and ArduPilot, identifying eight UAV-specific root-cause categories and five challenges in detection and reproduction tied to the cyber-physical nature of these systems.
 
\noindent{\bf Position of This Work.} MAVLink protocol research focuses on malformed packets and unauthorized injection. Cryptographic solutions assume that authentication resolves security concerns, but authenticated commands can still exploit control system dynamics. Physical sensor research bypasses software but requires specialized equipment and proximity. \ac{IDS} detect protocol anomalies but may not catch attacks where there is no pattern of commands or malformed commands. Our hypothesis is that a properly authenticated command can still be malicious if the values it carries push the control system beyond stability boundaries and cause unintended behavior. This paper investigates this class of attacks.
 
Compared to Marty~\cite{marty2014vulnerability} and Kwon et al.~\cite{kwon2018empirical}, our attacks use only well-formed, protocol-compliant commands. Compared to Son et al.~\cite{son2015rocking} and Trippel et al.~\cite{trippel2017walnut}, our attacks need no physical proximity. Compared to Shen et al.~\cite{shen2020drift}, who exploit Kalman-filter uncertainty windows through GPS spoofing, we show that a single \texttt{PARAM\_SET} message inflating EKF noise covariances produces the same kind of filter divergence on a UAV without external signal injection. Compared to Erba et al.~\cite{erba2024sensor}, our attacks target the autopilot parameter store rather than sensor configuration registers, and the parameter store is reachable from any authenticated MAVLink client.

\begin{table}[tbh]
\centering
\caption{Summary of evaluated control-aware attacks along with the target controller and outcomes.}
\label{tab:attack_summary}
\begin{tabular}{p{.02in}p{1.2in}p{1.1in}p{.5in}}
\toprule
\textbf{\#} & \textbf{Attack} & \textbf{Target layer} & \textbf{Outcome} \\
\midrule
1 & Oscillation Injection         & Position controller            & Partial \\
2 & PID Gain Manipulation         & Pitch rate controller          & Partial \\
3 & EKF Confusion                 & State estimation               & Partial \\
4 & Bypassing Failsafe Constraints & Rate controller + position     & Dange-rous \\
5 & Yaw Axis Destabilization      & Yaw rate controller            & Partial \\
6 & Combined Multi-Layer          & EKF + rate + attitude          & Crash \\
\bottomrule
\end{tabular}
\end{table}

\begin{table*}[!h]
\centering
\caption{Parameters modified by the control-aware attacks in this work. Source: ArduPilot Copter parameter documentation~\cite{ardupilot_copter_params} and EKF3 source code~\cite{ardupilot_ekf3_source}.}
\label{tab:params_combined}
\begin{tabularx}{\textwidth}{l l X}
\toprule
\textbf{Parameter} & \textbf{Subsystem} & \textbf{Description} \\
\midrule
\multicolumn{3}{l}{\textbf{Rate controller parameters}} \\\midrule
\texttt{ATC\_RAT\_PIT\_P}    & Pitch rate PID & Proportional gain \\
\texttt{ATC\_RAT\_PIT\_I}    & Pitch rate PID & Integral gain \\
\texttt{ATC\_RAT\_PIT\_D}    & Pitch rate PID & Derivative gain (damping) \\
\texttt{ATC\_RAT\_PIT\_IMAX} & Pitch rate PID & Maximum integral accumulation \\
\texttt{ATC\_RAT\_RLL\_I}    & Roll rate PID  & Integral gain \\
\texttt{ATC\_RAT\_RLL\_D}    & Roll rate PID  & Derivative gain (damping) \\
\texttt{ATC\_RAT\_RLL\_IMAX} & Roll rate PID  & Maximum integral accumulation \\
\texttt{ATC\_RAT\_YAW\_P}    & Yaw rate PID   & Proportional gain \\
\texttt{ATC\_RAT\_YAW\_I}    & Yaw rate PID   & Integral gain \\
\texttt{ATC\_RAT\_YAW\_D}    & Yaw rate PID   & Derivative gain (damping) \\
\texttt{ATC\_RAT\_YAW\_IMAX} & Yaw rate PID   & Maximum integral accumulation \\
\addlinespace\midrule
\multicolumn{3}{l}{\textbf{Attitude controller parameters}} \\\midrule
\texttt{ATC\_ANG\_PIT\_P}    & Pitch attitude  & Proportional gain for pitch angle correction \\
\texttt{ATC\_ANG\_RLL\_P}    & Roll attitude   & Proportional gain for roll angle correction \\
\texttt{ATC\_ANG\_YAW\_P}    & Yaw attitude    & Proportional gain for yaw angle correction \\
\texttt{ATC\_SLEW\_YAW}      & Yaw controller  & Maximum yaw target slew rate (cdeg/s) \\
\addlinespace\midrule
\multicolumn{3}{l}{\textbf{EKF3 process, bias, and measurement-noise parameters}} \\\midrule
\texttt{EK3\_ACC\_P\_NSE}    & Process noise   & Accelerometer noise (m/s\textsuperscript{3}); higher = less trust in IMU prediction \\
\texttt{EK3\_GYRO\_P\_NSE}   & Process noise   & Gyroscope noise (rad/s); higher = less trust in angular-rate prediction \\
\texttt{EK3\_ABIAS\_P\_NSE}  & Bias estimate   & Accelerometer bias drift rate; higher = faster bias wander \\
\texttt{EK3\_GBIAS\_P\_NSE}  & Bias estimate   & Gyroscope bias drift rate; higher = faster bias wander \\
\texttt{EK3\_POSNE\_M\_NSE}  & Meas. noise     & Horizontal position noise (m); higher = weaker GPS correction \\
\texttt{EK3\_ALT\_M\_NSE}    & Meas. noise     & Altitude measurement noise; higher = weaker barometer correction \\
\texttt{EK3\_VELD\_M\_NSE}   & Meas. noise     & Vertical velocity noise; higher = weaker vertical correction \\
\addlinespace\midrule
\multicolumn{3}{l}{\textbf{EKF3 gates, safety checks, and source configuration}} \\\midrule
\texttt{EK3\_POS\_I\_GATE}   & Innovation gate & Position outlier rejection threshold; higher = accepts worse measurements \\
\texttt{EK3\_HGT\_I\_GATE}   & Innovation gate & Height outlier rejection threshold; higher = accepts worse measurements \\
\texttt{EK3\_GPS\_CHECK}     & EKF safety      & Bitmask for GPS health checks; 0 = all checks disabled \\
\texttt{EK3\_GLITCH\_RAD}    & EKF safety      & GPS glitch detection radius (m); 0 = disabled \\
\texttt{EK3\_SRC1\_POSXY}    & Source config   & Horizontal position source; 0 = none \\
\texttt{EK3\_SRC1\_VELXY}    & Source config   & Horizontal velocity source; 0 = none \\
\texttt{EK3\_SRC1\_POSZ}     & Source config   & Vertical position source; 0 = none \\
\addlinespace\midrule
\multicolumn{3}{l}{\textbf{Failsafe parameters}} \\\midrule
\texttt{FS\_EKF\_THRESH}     & EKF failsafe    & Variance threshold that triggers EKF failsafe \\
\bottomrule
\end{tabularx}
\end{table*}

\label{sec:simulation}
 
\section{Attack Design and Simulation}

This section presents our six control-aware attacks against ArduPilot. Each attack targets a different subsystem or combination of subsystems in the control hierarchy described in Section~\ref{sec:controllers}. All attacks, listed in Table~\ref{tab:attack_summary}, use only legitimate \ac{MAVLink} commands. ArduPilot accepts them because they are well formed and use the protocol as designed; their malicious effects arise from the selected values and from the timing and frequency with which the commands are sent.
 
\subsection{Simulation Setup}
 
We developed and evaluated the attacks using ArduPilot's Software-in-the-Loop (SITL) simulator. SITL compiles the actual ArduPilot firmware as an executable and replaces physical sensors and actuators with a flight-dynamics model. 
For each attack, we armed the vehicle, had it take off to a stable hover, and then ran the attack script while logging telemetry. The scripts connect to ArduPilot via \texttt{pymavlink}~\cite{mavlink} on \texttt{tcp:127.0.0.1:5763} and use the MAVLink messages described in Section~\ref{sec:controllers}. Table~\ref{tab:params_combined} describes the parameters modified by the attacks. 

All degradation percentages reported in this section were computed offline from the ArduPilot's DataFlash log for each flight. For each attack, we first recorded a baseline log of the mission, from the starting position through return-to-home, with no attack script running. We then recorded a log of the same mission with the attack script activated. A script parsed both logs, extracted the same telemetry metrics, and reported the relative difference between the attack run and the baseline scenario over the attack window. We report one representative experiment for each case; repeated exploratory trials showed similar behavior, but we did not collect enough repetitions to report means and standard deviations.

All scripts and logs are available on our GitHub page~\cite{AttackScripts2026}.
    
\subsection{Attack 1: Oscillation Injection}
 
This attack induces oscillatory behavior by alternating position setpoints sent through the MAVLink command \texttt{SET\_POSITION\_TARGET\_LOCAL\_NED} when the drone is in Guided mode. We command the vehicle to move approximately 20~m north and south at 10~Hz for 30~seconds. The position controller computes a velocity target proportional to the position error at the default frequency of 50~Hz. When we flip the target to the opposite side, the error sign reverses and the controller commands velocity in the new direction, but the vehicle still has inertia toward the old target. This creates a large velocity error.
 
The attack tests whether the ArduPilot can reject a sequence of well-formed but adversarially timed targets; i.e., this attack does not modify any controller parameters. In the tested configuration, we observe that ArduPilot did not perform this type of validation. 
 
\noindent\textbf{Results.} We observed that the pitch-rate variability rose by 133\%, with peak rates of 40.3$^{\circ}$/s, and the roll-rate variability rose by 186\%. The attitude oscillations stayed below the 15$^{\circ}$ failsafe threshold, so no failsafe was triggered -- See Section~\ref{sec:failsafes} for details about the implemented fail-safe mechanisms. However, the altitude tracking degraded, the position trajectory deviated by tens of meters from the commanded path, and the GPS ground track showed clear oscillation. During this attack, the vehicle remained airborne but exhibited visibly degraded flight.
 
\subsection{Attack 2: PID Gain Manipulation}
 
The pitch-rate controller is the innermost feedback loop for the pitch axis. It runs at 400~Hz with default gains \texttt{ATC\_RAT\_PIT\_P}~$\approx 0.135$, \texttt{ATC\_RAT\_PIT\_I}~$\approx 0.135$, \texttt{ATC\_RAT\_PIT\_D}~$\approx 0.0036$, and \texttt{ATC\_RAT\_PIT\_IMAX}~$\approx 0.5$; see Table~\ref{tab:params_combined} for descriptions of these parameters. We modify these parameters via the MAVLink command \texttt{PARAM\_SET} to \texttt{ATC\_RAT\_PIT\_I}~$= 0.500$ (a 3.7$\times$ increase, causing integral windup), \texttt{ATC\_RAT\_PIT\_D}~$= 0.0001$ (a 36$\times$ decrease, causing near-zero damping), and \texttt{ATC\_RAT\_PIT\_IMAX}~$= 1.0$ (doubled). We leave the roll and yaw axes at their default values.
 
\noindent\textbf{Results.} Pitch-rate variability rose by 3\%, and the attitude trace showed minor irregularities. However, the vehicle continued to fly normally, and no failsafe was triggered. The asymmetric corruption, which targeted pitch only, was not enough on its own to push the controller across the stability boundary. The important finding is not the limited visible effect, but that ArduPilot accepts and applies these gain values in real time without control-aware checking, operator confirmation, or rate limiting. This weakness is exploited in Attacks~4 and~6.

\subsection{Attack 3: \acf{EKF} Confusion}
\label{sec:sim_attack3}

Attack~3 targets the state estimator rather than the attitude loops directly. Under normal operation, EKF3 balances short-term IMU prediction against GPS, barometer, and magnetometer corrections. The default parameter values used in our experiments are \texttt{EK3\_ACC\_P\_NSE}$\approx0.35$, \texttt{EK3\_GYRO\_P\_NSE}$\approx0.015$, \texttt{EK3\_POSNE\_M\_NSE}$\approx0.5$, \texttt{EK3\_POS\_I\_GATE}$\approx5$, and \texttt{EK3\_GPS\_CHECK}$=31$; see Table~\ref{tab:params_combined} for descriptions of these parameters. These parameters control the estimator's trust model: process noise affects prediction covariance, measurement noise affects correction strength, and gates/checks reject implausible sensor updates.

The attack corrupts nine \ac{EKF} parameters in three phases. First, process noise is inflated so the filter distrusts its own inertial prediction. This is achieved by setting \texttt{EK3\_ACC\_P\_NSE}$=5.0$, \texttt{EK3\_GYRO\_P\_NSE}$=3.0$, \texttt{EK3\_ABIAS\_P\_NSE}$=1.0$, and \texttt{EK3\_GBIAS\_P\_NSE}$=0.5$. Second, measurement noise is inflated so GPS, barometer, and vertical-velocity corrections carry less weight. This is achieved by setting \texttt{EK3\_POSNE\_M\_NSE}$=10.0$, \texttt{EK3\_ALT\_M\_NSE}$=5.0$, and \texttt{EK3\_VELD\_M\_NSE}$=5.0$. Third, the safety envelope is widened by setting \texttt{EK3\_POS\_I\_GATE}$=1000$ and \texttt{EK3\_GPS\_CHECK}$=0$. Each value is sent through MAVLink command as \texttt{PARAM\_SET} and is applied immediately.

In the EKF prediction step, inflated process noise increases-- see Equation~\ref{eq:ekf_predict}. In the update step, shown in Equation~\ref{eq:ekf_update}, inflated measurement noise increases the denominator of the Kalman gain, causing sensor corrections to be underweighted. The bias-noise parameters let accelerometer and gyro bias estimates wander, and the disabled GPS checks remove the protection that would normally reject implausible updates. Downstream controllers then operate on corrupted state estimates without knowing that the estimator has been manipulated.

\noindent\textbf{Why this works against the EKF.} The attack exploits the structural trust the filter places in its own configuration. ArduPilot's EKF3 has no notion of a ``reasonable'' noise value: the parameters $Q_t$ (process noise) and $R_t$ (measurement noise) are treated as ground truth. By inflating $Q_t$, the attacker tells the filter that its inertial prediction is unreliable, so it leans harder on sensor corrections. By simultaneously inflating $R_t$, the attacker tells the filter that the sensors are also unreliable, so the Kalman gain $K_t$ shrinks toward zero and corrections are effectively ignored. With both inflated, the filter trusts nothing: it neither predicts well nor corrects well, and the state estimate drifts along the accumulated bias. Crucially, the filter's own covariance bound $P_{t|t-1}$ stays within the range Ardupilot considers healthy because the inflated noise values make the larger covariance look expected rather than anomalous. This is why \texttt{FS\_EKF\_THRESH} does not trigger the fail-safe mechanisms, which compare the innovation variance against the threshold. 

\begin{figure}[htb]
\centering
  \includegraphics[width=.95\linewidth]{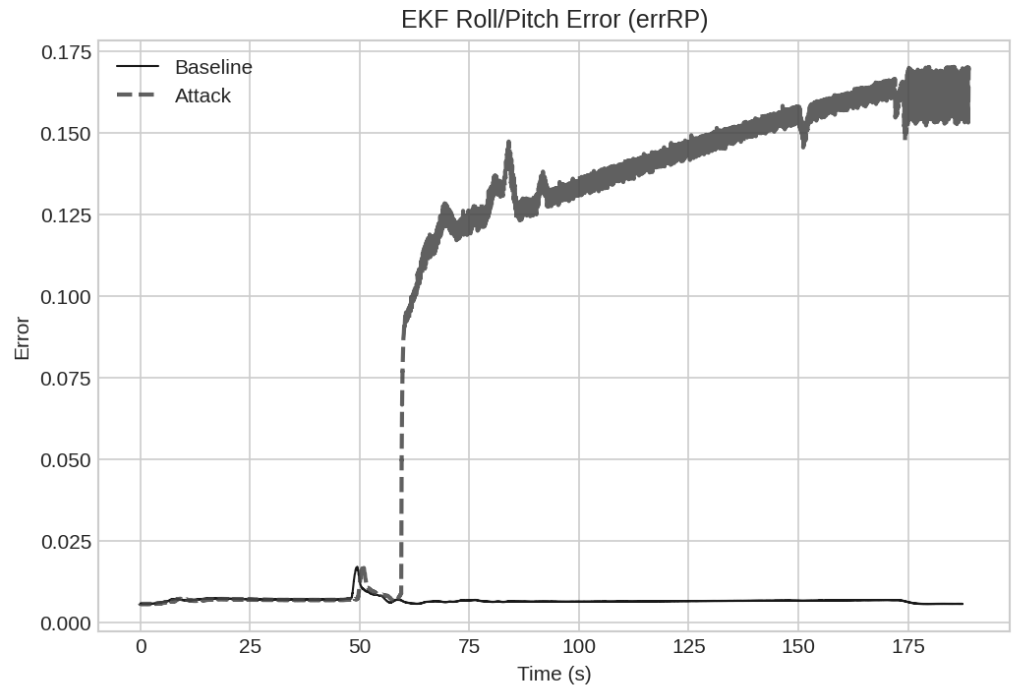}
  \caption{Attack 3 - The EKF roll/pitch error \texttt{errRP}. Note that the error is almost 0 in the baseline scenario and ramps to 0.170 while the vehicle remains airborne in the attack scenario.}
\label{fig:attack3-errrp}
\end{figure}

\noindent\textbf{Simulation Results.} Figure~\ref{fig:attack3-errrp} shows the error traces during the attack simulation. This attack damages the state estimator where the EKF roll/pitch innovation magnitude \texttt{errRP} rises from a baseline mean of 0.007 to a mean of 0.100, with a peak of 0.170. This corresponds to an increase of roughly 1{,}376\% relative to baseline. The ramp begins near $t=60$~s, which is when the inflated measurement noise has reduced the Kalman gain enough that the filter stops correcting accumulated prediction error.

An operator watching the attitude and rate panels from the authorized GCS would not see the effect of this attack. The external flight behavior remained almost unchanged -- roll, pitch, and yaw variability changed by less than 1\%, and angular rates shifted by less than 0.3\%. 
Even though the attack does not crash the drone or activate the failsafe mechanism, 
instead it establishes internal damage in the controllers via a corrupted but failsafe-silent state estimate. In Attack~6, we exploit this weakness by combining it with rate-loop corruption so that the silently incorrect estimate is fed into deliberately destabilized controllers. Thus, monitoring attitude and rate alone is insufficient to detect this attack. Independent monitoring of \ac{EKF} innovation health is required to \textit{mitigate} this attack.

\subsection{Attack 4: Bypassing Failsafe Constraints}
\label{sec:sim_attack4} 
The attacker corrupts both the pitch and roll rate loops by issuing commands that set \texttt{ATC\_RAT\_PIT\_I} to 0.5 and \texttt{ATC\_RAT\_RLL\_I} to 0.8, set \texttt{ATC\_RAT\_PIT\_D}=\texttt{ATC\_RAT\_RLL\_D} to 0.0, and raise both \texttt{IMAX} values to 1.5--affecting the altitude and the rate controller of Eq.~\ref{eq:rate_pid}. The attacker then injects circular position oscillations that repeatedly excite the weakened inner loops. The script sends a three-dimensional oscillating position target in which latitude and longitude trace a circular pattern while altitude oscillates independently.


\begin{figure}[tb]
\centering
\begin{subfigure}[t]{0.48\textwidth}\centering
  \includegraphics[width=\linewidth]{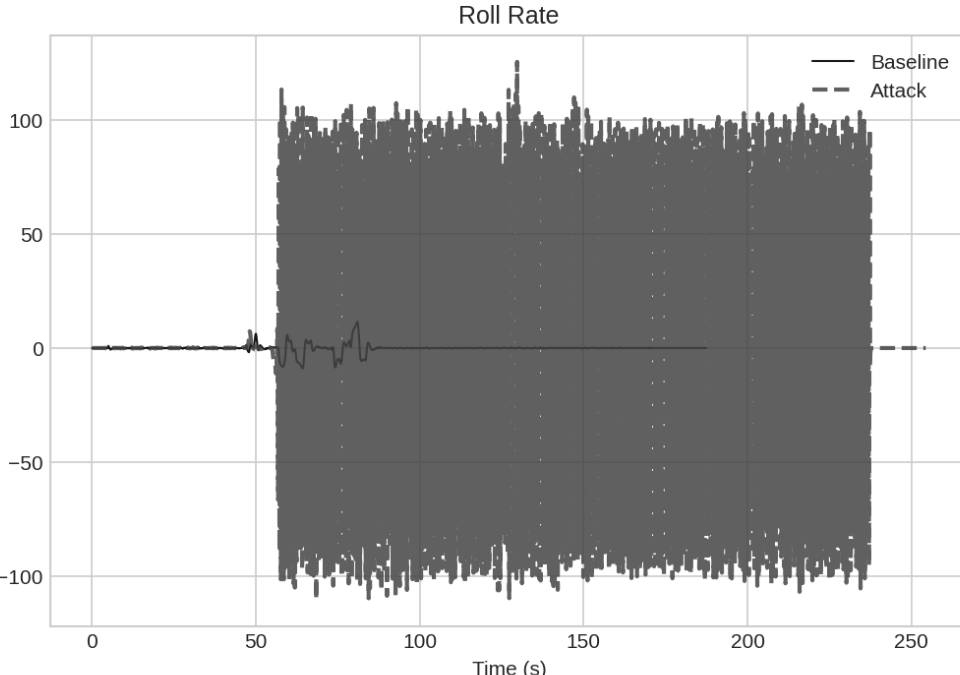}
  \caption{Roll rate.}
  \label{fig:attack4-rollrate}
\end{subfigure}
\begin{subfigure}[t]{0.48\textwidth}\centering
  \includegraphics[width=\linewidth]{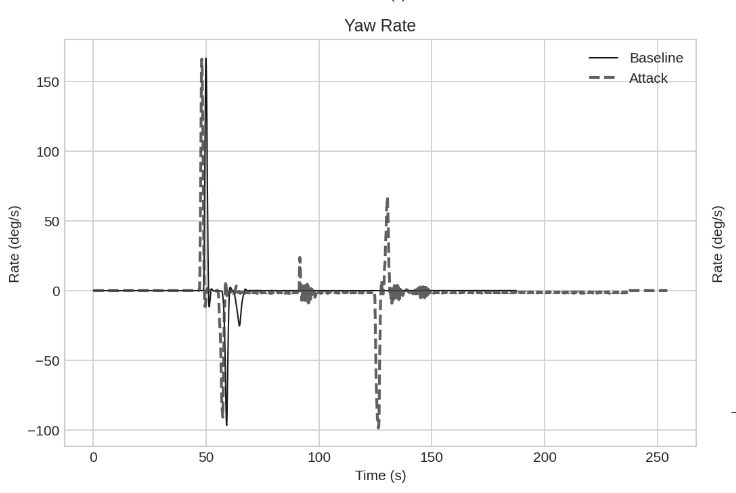}
  \caption{Yaw rate.}
  \label{fig:attack4-yawrate}
\end{subfigure}
\caption{Attack~4 angular-rate response. (a)~Roll rate enters sustained high-amplitude oscillation between roughly $\pm100^{\circ}$/s once the PID corruption and setpoint excitation interact. (b)~Yaw rate stays close to baseline most of the time but shows large transient spikes near the setpoint-flip events, exposing the cross-axis coupling.}
\label{fig:attack4-rates}
\end{figure}

The corrupted integral terms accumulate error quickly while zero derivative damping removes the term that normally suppresses oscillation. The rotating position target prevents the outer loop from converging: Every time the cascade begins correcting toward one target, the target moves and because both roll and pitch are corrupted, the vehicle loses stable lateral axis. The altitude oscillation stresses thrust compensation, i.e., when the vehicle leans, the vertical thrust component decreases, and the altitude controller demands more thrust and pushes the motors closer to saturation.

\noindent\textbf{Results.} Attack~4 severely degraded ArduPilot control by increasing the roll-rate variability by 3{,}155\%, with peaks near 320$^{\circ}$/s in SITL. The SITL's DataFlash analysis also shows sustained $\pm100$--$125^{\circ}$/s oscillation after the attack onset. Pitch-rate variability increased by 2{,}533\%, tracking errors increased by 437\% in roll and 573\% in pitch, and \texttt{errRP} jumped to approximately 0.15. The vehicle drifted southward and remained airborne in a severely degraded state rather than failing cleanly. 

\begin{figure}[tb]
\centering
\begin{subfigure}[t]{0.48\textwidth}\centering
  \includegraphics[width=\linewidth]{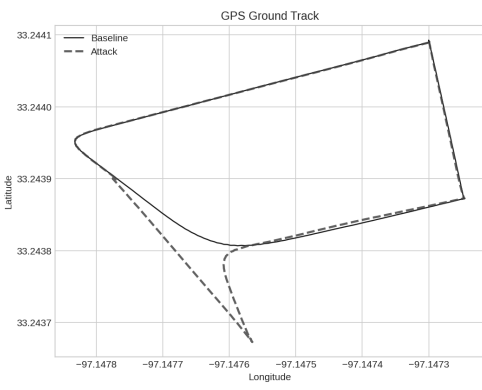}
  \caption{GPS ground track.}
  \label{fig:attack4-gps}
\end{subfigure}\hfill
\begin{subfigure}[t]{0.48\textwidth}\centering
  \includegraphics[width=\linewidth]{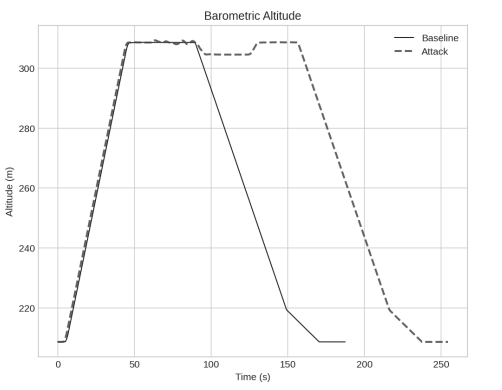}
  \caption{Barometric altitude.}
  \label{fig:attack4-baro}
\end{subfigure}
\caption{Attack~4 trajectory response. The ground track deviates significantly from baseline, and barometric altitude drifts as the destabilized inner loops corrupt thrust compensation.}
\label{fig:attack4-trajectory}
\end{figure}

Figure~\ref{fig:attack4-rates} shows the angular-rate response on the roll and yaw axes. We observe that, while the yaw rate changes instantly, the roll rate oscillates between extreme values for about 3 minutes. Figure~\ref{fig:attack4-trajectory} shows the ArduPilot deviations in terms of ground track and barometric-altitude deviation. Figure~\ref{fig:attack4-gps} shows minor deviations in latitude and longitude, while Figure~\ref{fig:attack4-baro} shows a delayed landing. Together, these show how the destabilized inner loops degrade the altitude and trajectory tracking simultaneously while the vehicle remains airborne.

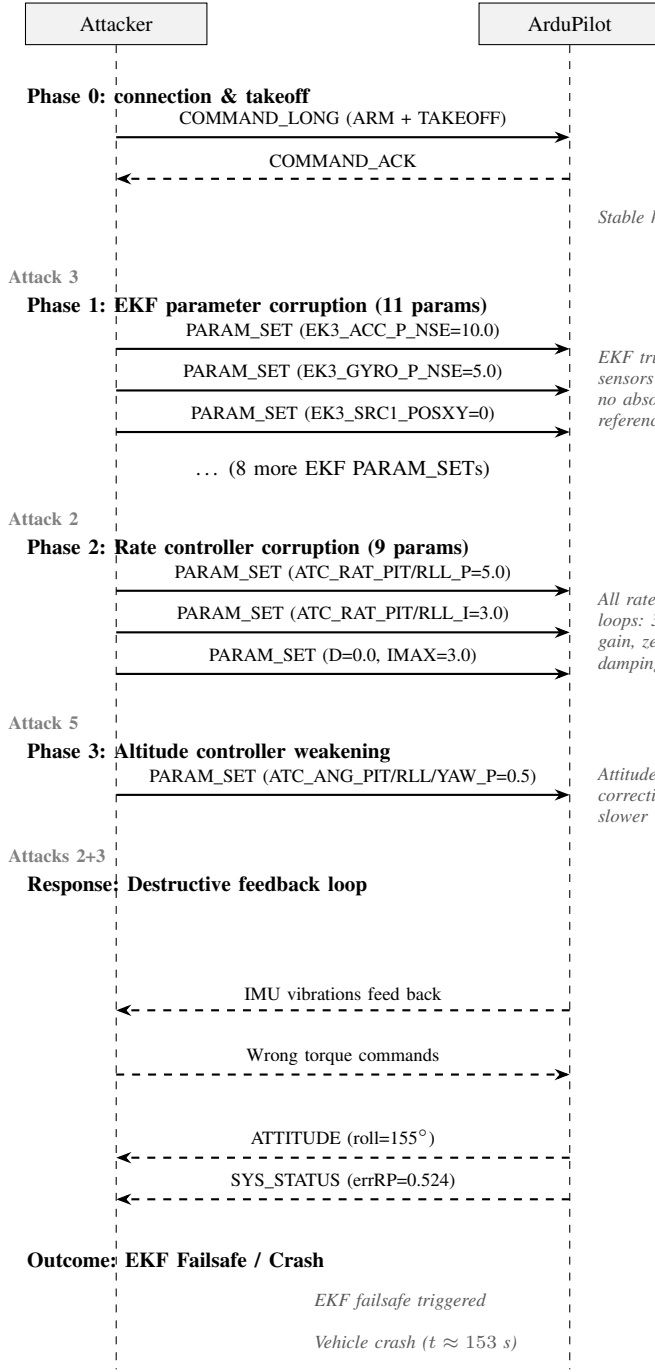
\begin{figure}[!htb]
\centering
\begin{tikzpicture}[
  font=\footnotesize,
  msg/.style={-{Stealth[length=2mm]},thick},
  ret/.style={-{Stealth[length=2mm]},thick,dashed},
  note/.style={font=\scriptsize\itshape,text=black!65,align=left},
  phase/.style={font=\footnotesize\bfseries},
  src/.style={font=\scriptsize\bfseries,text=black!50},
  x=1cm,y=1cm
]

\def\A{0.8}
\def\P{6.8}

\node[draw,fill=black!5,minimum width=2.4cm,minimum height=0.55cm] (atk) at (\A,0)
{ Attacker};

\node[draw,fill=black!5,minimum width=2.4cm,minimum height=0.55cm] (ap) at (\P,0)
{ArduPilot};


\def\bottom{-17.8}

\draw[dashed] (atk.south) -- (\A,\bottom);
\draw[dashed] (ap.south) -- (\P,\bottom);

\node[phase,anchor=west] at (\A-1.3,-0.95)
{Phase 0: connection \& takeoff};

\draw[msg] (\A,-1.5) --
node[above,font=\scriptsize]
{COMMAND\_LONG (ARM + TAKEOFF)}
(\P,-1.5);

\draw[ret] (\P,-2.05) --
node[above,font=\scriptsize]
{COMMAND\_ACK}
(\A,-2.05);

\node[note,anchor=west] at (\P+.25,-2.55)
{Stable hover};

\node[src,anchor=west] at (\A-1.55,-3.35)
{Attack 3};

\node[phase,anchor=west] at (\A-1.3,-3.75)
{Phase 1: EKF parameter corruption (11 params)};

\draw[msg] (\A,-4.3) --
node[above,font=\scriptsize]
{PARAM\_SET (EK3\_ACC\_P\_NSE=10.0)}
(\P,-4.3);

\draw[msg] (\A,-4.85) --
node[above,font=\scriptsize]
{PARAM\_SET (EK3\_GYRO\_P\_NSE=5.0)}
(\P,-4.85);

\draw[msg] (\A,-5.4) --
node[above,font=\scriptsize]
{PARAM\_SET (EK3\_SRC1\_POSXY=0)}
(\P,-5.4);

\node at (\A+3,-5.9)
{\dots\ (8 more EKF PARAM\_SETs)};

\node[note,anchor=west] at (\P+0.25,-4.85)
{EKF trusts \\sensors less; \\no absolute \\reference};

\node[src,anchor=west] at (\A-1.55,-6.55)
{Attack 2};

\node[phase,anchor=west] at (\A-1.3,-6.95)
{Phase 2: Rate controller corruption (9 params)};

\draw[msg] (\A,-7.5) --
node[above,font=\scriptsize]
{PARAM\_SET (ATC\_RAT\_PIT/RLL\_P=5.0)}
(\P,-7.5);

\draw[msg] (\A,-8.05) --
node[above,font=\scriptsize]
{PARAM\_SET (ATC\_RAT\_PIT/RLL\_I=3.0)}
(\P,-8.05);

\draw[msg] (\A,-8.6) --
node[above,font=\scriptsize]
{PARAM\_SET (D=0.0, IMAX=3.0)}
(\P,-8.6);

\node[note,anchor=west] at (\P+0.25,-8.05)
{All rate \\loops: 37x \\gain, zero \\damping};

\node[src,anchor=west] at (\A-1.55,-9.25)
{Attack 5};

\node[phase,anchor=west] at (\A-1.3,-9.65)
{Phase 3: Altitude controller weakening};

\draw[msg] (\A,-10.2) --
node[above,font=\scriptsize]
{PARAM\_SET (ATC\_ANG\_PIT/RLL/YAW\_P=0.5)}
(\P,-10.2);

\node[note,anchor=west] at (\P+0.25,-10.2)
{Attitude \\correction 9x \\slower};

\node[src,anchor=west] at (\A-1.55,-11.0)
{Attacks 2+3};

\node[phase,anchor=west] at (\A-1.3,-11.4)
{Response: Destructive feedback loop};

\draw[ret] (\P,-13.05) --
node[above,font=\scriptsize]
{IMU vibrations feed back}
(\A,-13.05);

\draw[ret] (\A,-13.9) --
node[above,font=\scriptsize]
{Wrong torque commands}
(\P,-13.9);

\draw[ret] (\P,-15.0) --
node[above,font=\scriptsize]
{ATTITUDE (roll=155$^\circ$)}
(\A,-15.0);

\draw[ret] (\P,-15.55) --
node[above,font=\scriptsize]
{SYS\_STATUS (errRP=0.524)}
(\A,-15.55);

\node[phase,anchor=west] at (\A-1.3,-16.35)
{Outcome: EKF Failsafe / Crash};

\node[note,anchor=west] at (\P-3.5,-16.9)
{EKF failsafe triggered};

\node[note,anchor=west] at (\P-3.5,-17.45)
{Vehicle crash ($t\approx153$ s)};

\end{tikzpicture}

\caption{UML Sequence Diagram for Attack 6}
\label{fig:attack6-seq}

\end{figure}

\begin{figure}[tb]
\centering
\begin{subfigure}[t]{0.48\textwidth}\centering
  \includegraphics[width=\linewidth]{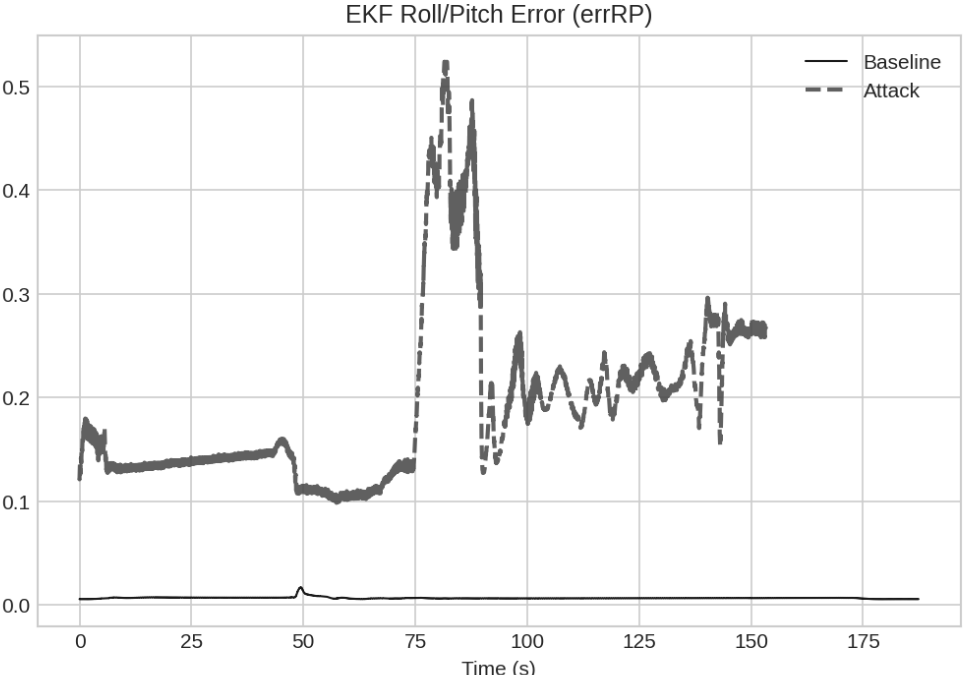}
  \caption{EKF roll/pitch error \texttt{errRP}.}
  \label{fig:attack6-errrp}
\end{subfigure}

\begin{subfigure}[t]{0.48\textwidth}\centering
  \includegraphics[width=\linewidth]{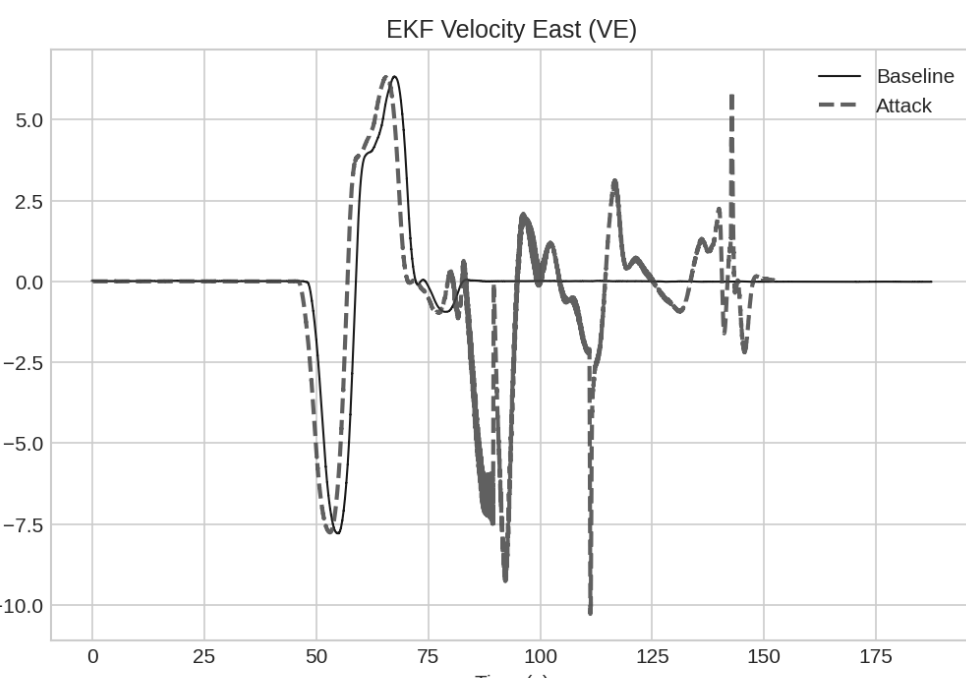}
  \caption{Velocity east (VE).}
  \label{fig:attack6-ve}
\end{subfigure}
\caption{Attack~6 estimator response. \texttt{errRP} peaks at 0.524 before the EKF failsafe triggers and the north/east velocity estimates show the filter losing coherent state long before the failsafe activates.}
\label{fig:attack6-ekf}
\end{figure}

\begin{figure}[tb]
\centering
\begin{subfigure}[t]{0.48\textwidth}\centering
  \includegraphics[width=\linewidth]{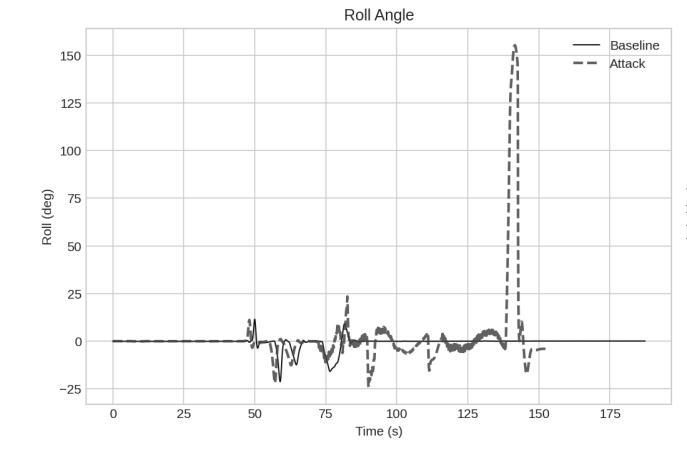}
  \caption{Roll angle.}
  \label{fig:attack6-roll}
\end{subfigure}\hfill
\begin{subfigure}[t]{0.48\textwidth}\centering
  \includegraphics[width=\linewidth]{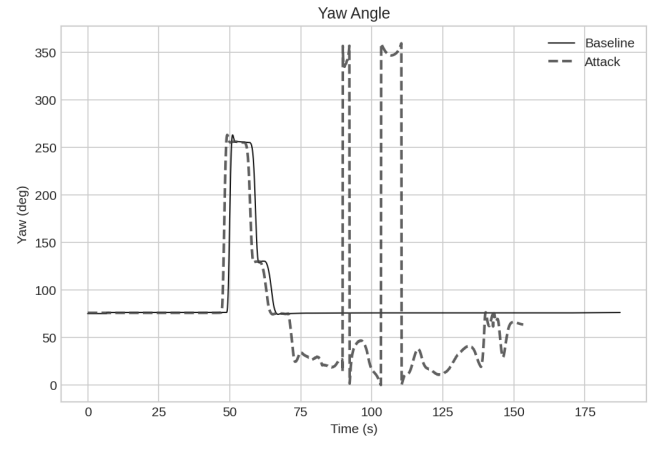}
  \caption{Yaw angle.}
  \label{fig:attack6-yaw}
\end{subfigure}
\caption{Attack~6 attitude response. Roll diverges to approximately $155^{\circ}$ (near inversion) and yaw drifts uncontrollably as the corrupted controllers and corrupted estimator drive each other.}
\label{fig:attack6-attitude}
\end{figure}

\subsection{Attack 5: Yaw Axis Destabilization}
 
The yaw controller maintains heading orientation. It is normally limited by \texttt{ATC\_SLEW\_YAW} ($\sim$6{,}000~cdeg/s) and \texttt{RATE\_Y\_MAX}. We modify six yaw-related parameters: \texttt{ATC\_RAT\_YAW\_P}~$=1.5$ (8$\times$), \texttt{ATC\_RAT\_YAW\_I}~$=0.5$ (28$\times$), and \texttt{ATC\_RAT\_YAW\_D}~$=0.0$--see Eq.~\ref{eq:rate_pid}. We widen the slew rate to 36{,}000~cdeg/s (6$\times$, removing slew protection) and weaken \texttt{ATC\_ANG\_YAW\_P}~$=0.5$--see Eq.~\ref{eq:motor_mix}. We then send \texttt{CONDITION\_YAW} commands at $\pm720^{\circ}$/s, with CW/CCW reversals every 0.5~seconds.
 
\noindent\textbf{Results.} The attack forced the yaw rate to peak above $\pm200^{\circ}$/s and caused cross-axis roll/pitch variability to rise visibly because the differential thrust required for yaw saturated the mixer, which prioritizes roll/pitch over yaw. The attack did not affect the copter's altitude. However, it exposes a design trade-off in the mixer: heading accuracy is sacrificed to maintain roll/pitch stability under saturation. Attack~6 exploits this design trade-off.

\subsection{Attack 6: Multi-Layer Combined Attack}
\label{sec:sim_attack6} 

Attack~6 combines \ac{EKF} corruption from Attack~3, rate-loop destabilization from Attacks~2 and~4, and altitude-loop weakening. The attack creates a destructive feedback loop in which corrupted state estimates and overtuned controllers reinforce one another even after the attacker stops transmitting.

The attack sends approximately 30 \texttt{PARAM\_SET} commands, organized into three groups, in under five seconds. Figure~\ref{fig:attack6-seq} shows the sequence of messages sent by the attacker and their impacts. The attack has three phases. In phase~1, we corrupt the EKF noise, gates, checks, and source selection by assigning extreme values to \texttt{EK3\_ACC\_P\_NSE}, \texttt{EK3\_GYRO\_P\_NSE}, \texttt{EK3\_ABIAS\_P\_NSE}, \texttt{EK3\_GBIAS\_P\_NSE}, \texttt{EK3\_POSNE\_M\_NSE}, \texttt{EK3\_ALT\_M\_NSE}, \texttt{EK3\_VELD\_M\_NSE}, \texttt{EK3\_POS\_I\_GATE}, \texttt{EK3\_HGT\_I\_GATE}, \texttt{EK3\_GPS\_CHECK}, and \texttt{EK3\_GLITCH\_RAD}. It also removes EKF position and velocity sources using \texttt{EK3\_SRC1\_POSXY}=0, \texttt{EK3\_SRC1\_VELXY}=0, and \texttt{EK3\_SRC1\_POSZ}=0.
In phase~2, we corrupt the pitch, roll, and yaw rate controllers. For the pitch and roll controllers, the atatck script sets \texttt{ATC\_RAT\_PIT\_P} and \texttt{ATC\_RAT\_RLL\_P} to 5.0, \texttt{ATC\_RAT\_PIT\_I} and \texttt{ATC\_RAT\_RLL\_I} to 3.0, \texttt{ATC\_RAT\_PIT\_D} and \texttt{ATC\_RAT\_RLL\_D} to 0, \texttt{ATC\_RAT\_PIT\_IMAX} and \texttt{ATC\_RAT\_RLL\_IMAX} to 3.0. For the yaw, the script sets parameter \texttt{ATC\_RAT\_YAW\_P} to 1.5, \texttt{ATC\_RAT\_YAW\_I}  to 0.5, and \texttt{ATC\_RAT\_YAW\_IMAX} to 2.0. Phase~3 weakens the outer attitude controllers by setting \texttt{ATC\_ANG\_PIT\_P}, \texttt{ATC\_ANG\_RLL\_P}, and \texttt{ATC\_ANG\_YAW\_P} to 0.5. The script then stops issuing attack commands and monitors telemetry and attack impact.

We emphasize that the values for \texttt{PARAM\_SET} messages were chosen empirically through trial and errors. In this paper, we report a successful attack scenario. Future work will derive analytical bounds on the parameters needed to guarantee the desired impact. 

\noindent\textbf{ArduPilot response to the attack commands.} Attack~6 requires no attacker input after the initial parameter burst. The attack corrupts three input and command processing layers impacting the controller's feedback loop. The overtuned controllers, with controller gain constants $P$ and $I$, are raised by roughly $37\times$ and derivative damping set to zero, oscillate at high frequency, and that oscillation shakes the airframe. The IMU registers the vibration as a genuine motion. Normally, the \ac{EKF} would reject this using GPS, barometer, and velocity corrections. However, Phase~1 has set \texttt{EK3\_SRC1\_POSXY}, \texttt{EK3\_SRC1\_VELXY}, and \texttt{EK3\_SRC1\_POSZ} to zero and widened every gate, so the filter has no independent reference to contradict the vibration-corrupted inertial data. Therefore, the resulting incorrect state estimate is passed to the weakened attitude controllers, whose constant $P$ gains are reduced to roughly $1/9$ of their default values and are therefore too slow to correct the growing error. Their late and incorrect torque commands drive the rate loops harder, which increases vibration and further corrupts the estimation where with each iteration of the loop, errors are amplified. Therefore, the attacker does not need to continue transmitting after about five seconds as the vehicle still proceeds to failure on its own.

\noindent\textbf{Results.} Attack~6 forced Ardupilot for complete loss of control. Figure~\ref{fig:attack6-ekf} shows the roll/pitch error while Figure~\ref{fig:attack6-errrp} shows \texttt{errRP} reaching 0.524 and causing the east velocity of Figure~\ref{fig:attack6-ve} to become unstable. Figure~\ref{fig:attack6-attitude} shows the attitude deviation in response to the combined attack. We observe that the roll (Figure~\ref{fig:attack6-roll}) reaches approximately $155^{\circ}$, indicating near inversion and the yaw fluctuates without control (Figure~\ref{fig:attack6-yaw}). These deviations saturate the throttle; see Eq.~\ref{eq:motor_mix}. Hence, the EKF failsafe triggers at approximately $t=153$~s, right after the roughly five-second parameter-corruption burst has ended, confirming that the loop is self-sustaining rather than driven by continued attacker input.

\subsection{Summary of Simulation Results}
 
The six attacks span a spectrum from partial degradation to complete loss of control. Attacks~1, 2, 3, and~5 produce partial degradation through different mechanisms: Command timing alone in Attack~1, single-axis gain corruption in Attack~2, state-estimator corruption invisible to attitude monitoring in Attack~3, and yaw destabilization with cross-axis coupling in Attack~5. Attack~4 produces a dangerous but non-crashing outcome, and Attack~6 produces complete loss of control by combining mechanisms across multiple control layers. Across all six attacks, the tested configuration applies parameter changes and command sequences without control-aware validation of whether the resulting behavior remains safe, reasonable, or consistent with stable flight. 
\section{Proof-of-concept validation Using a Pixhawk 2.4.8 Flight Controller}
\label{sec:validation}
 
\begin{table*}[htb]
\centering
\small
\caption{Hardware validation results.}
\label{tab:hw_results}
\begin{tabularx}{\textwidth}{l X}
\toprule
\textbf{Attack} & \textbf{Hardware behavior} \\
\midrule
1: Oscillation Injection          & Visible attitude oscillation at 2~Hz; increased pitch-rate peaks; yaw drift; altitude unaffected; no failsafe. \\
2: PID Gain Manipulation          & Minimal observable flight impact; corrupted gains confirmed via parameter read-back; no failsafe. This confirms that the tested configuration accepts unsafe PID values without control-aware validation. \\
3: EKF Confusion                  & Externally normal flight; \texttt{errRP} climbed steadily; no failsafe. This was the stealthiest attack: attitude monitoring alone could not detect EKF parameter corruption. \\
4: Bypassing Failsafe Constraints & Sustained $\pm 100^{\circ}$/s roll and pitch oscillation; visible southward drift; severely degraded but airborne flight. Effects were more pronounced than in SITL due to real vibrations. \\
5: Yaw Axis Destabilization       & Yaw rate exceeded $\pm 200^{\circ}$/s; cross-axis roll/pitch variability increased; the mixer prioritized roll/pitch as expected; altitude was unaffected. \\
6: Combined Multi-Layer           & Roll reached $155^{\circ}$, nearly inverted; roll rate exceeded $-512^{\circ}$/s; \texttt{errRP} peaked at 0.524; the drone crashed at $t \approx 153$~s when the EKF failsafe triggered. \\
\bottomrule
\end{tabularx}
\end{table*}

\begin{figure}[tb]
\centering
\includegraphics[width=.45\textwidth]{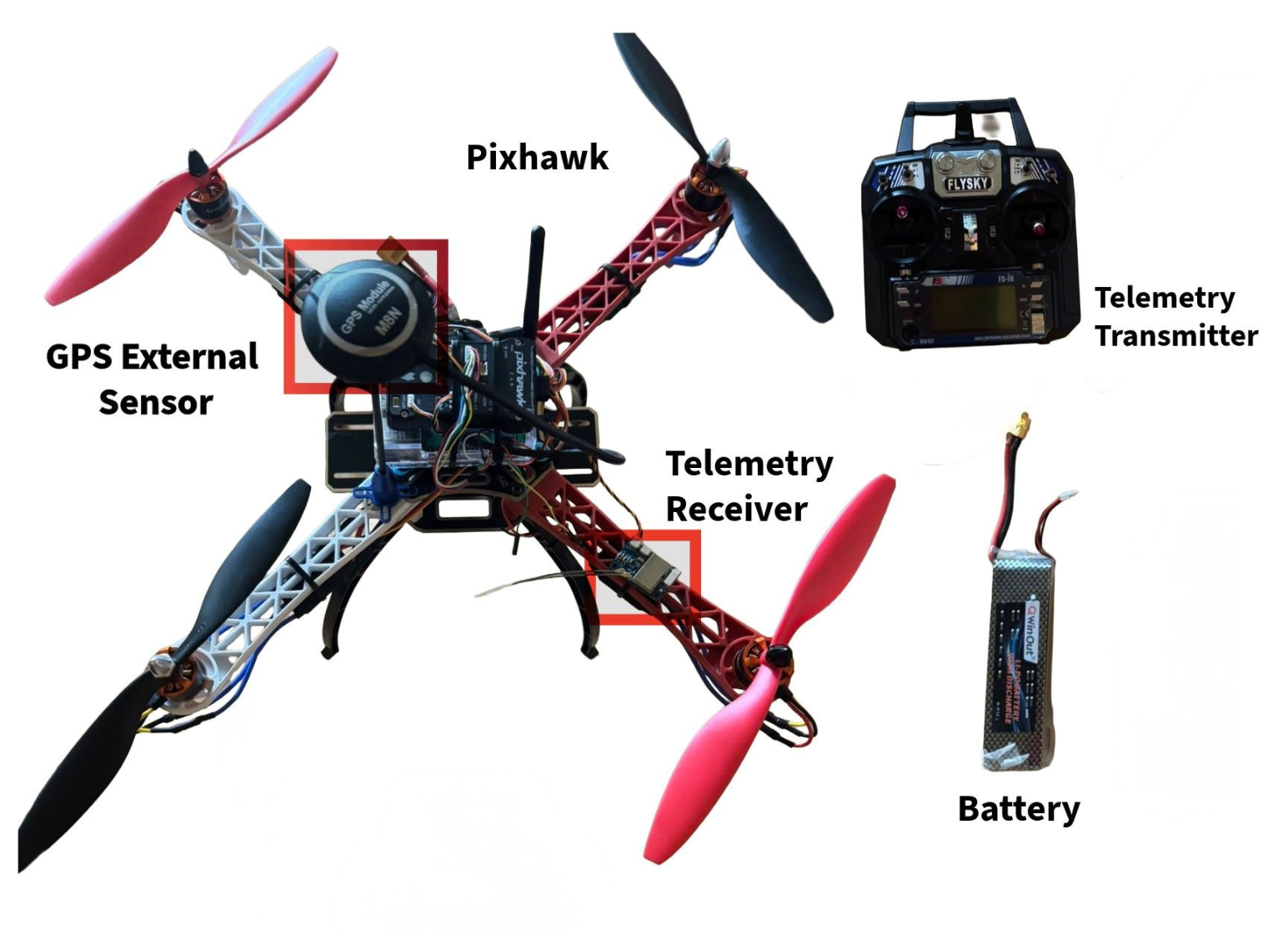}
\caption{Quadcopter testbed with a Pixhawk 2.4.8 flight controller used for hardware validation of the six attacks.}
\label{fig:hw_setup}
\end{figure}
\subsection{Hardware Setup}

To evaluate whether the SITL findings, we replicated the operations and attacks using real-world flight environment with a physical quadcopter (i.e., drone) under wind, motor noise, physical sensor behavior, and aerodynamic effects because although SITL runs the ArduCopter firmware, it does not fully capture these physical factors. 
The hardware experiments provide transfer evidence rather than exhaustive statistical validation. 
In the following, we first describe the hardware setup, then present and analyze the experimental results.

The physical testbed, shown in Figure~\ref{fig:hw_setup}, consisted of a Pixhawk 2.4.8 flight controller running ArduPilot Copter firmware, mounted on a quadcopter airframe with onboard GPS, a telemetry radio, a battery, and electronic speed controllers. The \acf{GCS} laptop ran Mission Planner~\cite{Obo2026} and the Python attack scripts~\cite{AttackScripts2026}. The attack scripts connect to the quadcopter over a 915~MHz SiK telemetry radio at 57{,}600~baud through a MAVProxy TCP splitter, which allows Mission Planner and the attack scripts to share the telemetry link. The scripts use \texttt{pymavlink} package to communicate with the quadcopter.
 
All flight experiments complied with U.S. Federal Aviation Administration regulations. The drone is registered under 14~CFR Part~48, registration number FA3TC7HCTX. Flights were conducted under the recreational operations exception, 49~U.S.C.~\S44809~\cite{FAA2029}.

\subsection{Experimental Results}

Each attack was executed at least once in flight, and ArduPilot's DataFlash logs were collected for post-flight analysis. Table~\ref{tab:hw_results} summarizes the observed behavior for each attack. Overall, the Pixhawk~2.4.8 hardware reproduced the same attack mechanisms observed in SITL: ArduPilot accepted and read back the tested \texttt{PARAM\_SET} commands without control-aware validation, and the drone exhibited the same categories of the degraded behavior documented in Section~\ref{sec:simulation}, including sustained attitude oscillation, EKF state drift, and controller instability. For Attack~6, the corruption window again lasted under five seconds, after which the drone entered a self-sustaining destructive feedback loop and crashed. In several cases, the hardware effects were more pronounced than in SITL because real airframe vibration and sensor noise provided additional excitation for the destabilized controllers to amplify.

ArduPilot's response to Attack~6 is of particular interest because it produced the most severe outcome. The corruption window of the attack lasted under five seconds. After the last \texttt{PARAM\_SET} command, no further attacker input was needed. The drone entered a self-sustaining destructive feedback loop. The 37$\times$ overtuned rate controllers oscillated at high frequency, producing airframe vibrations that the IMU registered as real motion. With \texttt{EK3\_SRC1\_POSXY}, \texttt{EK3\_SRC1\_VELXY}, and \texttt{EK3\_SRC1\_POSZ} all set to zero, the EKF could not distinguish these controller-induced vibrations from legitimate sensor data. The weakened attitude controllers, with gains reduced to roughly one-ninth of their default values, could not correct the growing attitude errors. The DataFlash logs confirm roll angles approaching $155^{\circ}$, roll rates exceeding $-512^{\circ}$/s, and an \texttt{errRP} peak of 0.524. The drone crashed approximately 153~seconds into the flight becuase the \texttt{errRP} exceeded the threshold \texttt{FS\_EKF\_THRESH}. These results show that ArduPilot's layered control architecture, although designed for modularity, can become a safety liability when an attacker corrupts multiple layers at once.

 \begin{figure*}[!tbh]
\centering
\includegraphics[width=1.0\textwidth]{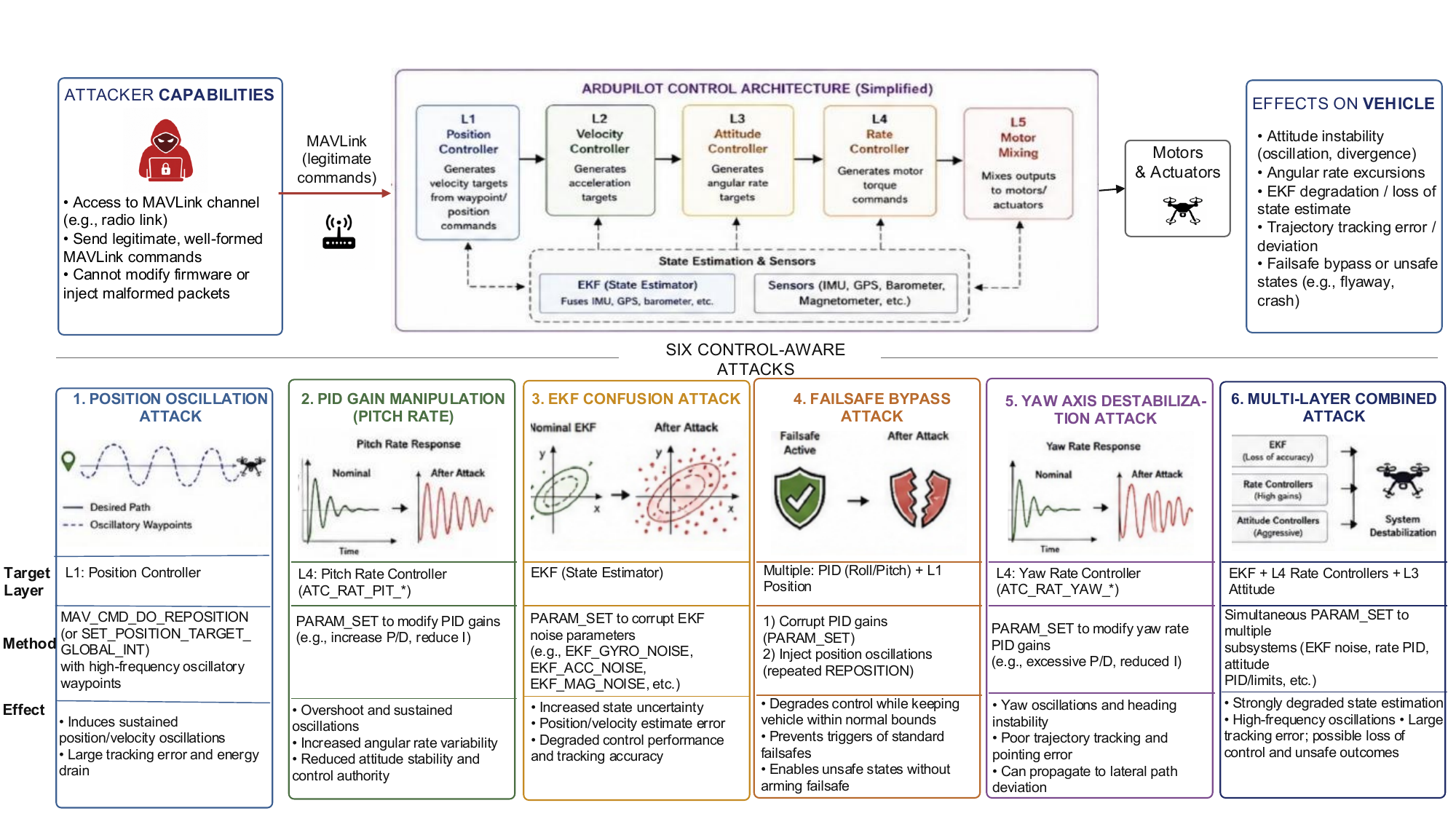}
\caption{Summary of the control-aware attacks, their methods, and their effects.}
\label{fig:summaryAttacks}
\end{figure*}

\begin{table*}[tbh]
\centering
\small
\caption{Mitigation impact across the six attacks.}
\label{tab:mitigation-impact}
\begin{tabularx}{\textwidth}{lccccccX}
\toprule
\textbf{Mitigation} & \textbf{A1} & \textbf{A2} & \textbf{A3} & \textbf{A4} & \textbf{A5} & \textbf{A6} & \textbf{Limitations} \\
\midrule
MAVLink signing          & Partial & Partial & Partial & Partial & Partial & Partial & Does not stop a compromised \acf{GCS}. \\
Parameter range checks   & No      & Yes     & Yes     & Yes     & Yes     & Yes     & Requires safe parameter ranges. \\
Parameter rate limiting  & No      & Partial & Partial & Partial & Partial & Yes     & Slows attacks but does not fully prevent them. \\
Control-aware monitoring & Yes     & Yes     & Yes     & Yes     & Yes     & Yes     & Requires model-specific thresholds. \\
\bottomrule
\end{tabularx}
\end{table*}

Several observations from the hardware experiments are worth noting. First, Ardupilot applies malicious parameters values without validation, including values that no legitimate tuning workflow would normally use, such as gyroscope process noise inflated by a factor of 200 and attitude proportional gain reduced ninefold. Second, the combined attack's self-sustaining property matters from a defensive perspective: The operator has roughly five seconds of anomalous parameter activity to detect and interrupt before the destructive loop closes. An operator monitoring Mission Planner's standard attitude and rate panels may not see malicious behavior while the navigation state silently degrades. Third, Attack~3 (i.e., the EKF Confusion) is the most concerning attack from an intrusion detection standpoint because it can remain externally quiet while preparing the estimator state that, when combined with controller corruption as in Attack~6, can contribute to a crash. 

\section{Discussion}
\label{sec:discussion}

\noindent\textbf{Summary.} Figure~\ref{fig:summaryAttacks} summarizes the six control-aware attacks against ArduPilot presented in this paper, in which an attacker uses only legitimate, well-formed MAVLink commands to drive the autopilot's control system into degraded or unsafe states. The attacks were evaluated in \ac{SITL} simulation and demonstrated on a Pixhawk 2.4.8 flight platform. The attacks span a spectrum from partial flight degradation (Attacks~1, 2, 3, and~5), through severely degraded flight (Attack~4), to complete loss of control (Attack~6). They exploit four distinct primitives that ArduPilot exposes to any authenticated MAVLink client: (1) command-timing manipulation, (2) runtime PID-gain modification, (3) EKF noise-covariance manipulation, and (4) cascade-control parameter corruption.
 
A consistent finding across all six attacks is that the tested configuration applies parameter changes and command sequences without control-aware validation of whether the resulting values are safe, reasonable, or consistent with stable flight. The autopilot's trust model assumes a benign operator, and that assumption creates exploitable gaps at every level of the cascaded control hierarchy.
 
\medskip
\noindent\textbf{Implications.} Securing the MAVLink communication channel is not enough. An attacker who compromises a legitimate \ac{GCS}, gains access through social engineering, or controls a malicious companion computer could damage the drone by manipulating the control system as presented in this paper. The commands resemble normal operations because they use standard interfaces, but they carry maliciously selected values and timing. Protocol-level \acf{IDS}~\cite{11140483,Tufekci2024DUDEIDS,Wu2024TinyUAVIDS,Ihekoronye2024DroneGuard} that focus on message format or rate patterns would also miss these attacks because the messages are well formed, use standard types, and arrive at expected rates. Effective detection would require an \ac{IDS} to understand the control-theoretic implications of command sequences and parameter values, which is more difficult than packet-pattern matching.
 
\medskip
\noindent\textbf{Mitigations and Future Work.} Table~\ref{tab:mitigation-impact} summarizes mitigation techniques that address the proposed attacks. Several directions follow naturally. \textit{Parameter validation} through range checking on \texttt{PARAM\_SET} could reject values exceeding a configurable multiple of the factory defaults. \textit{Rate limiting} on parameter changes would prevent the rapid multi-parameter corruption that enables Attack~6. Limiting modifications to one or two per second would extend the attack window from five seconds to minutes, giving operators time to notice the anomaly and giving control monitors time to respond. \textit{Cross-layer consistency monitoring} could flag the simultaneous modification of \ac{EKF}, rate, and attitude parameters within a short window, since legitimate tuning workflows do not modify these layers together. \textit{\ac{EKF} health monitoring} should be included in any \ac{IDS}, since attitude-only monitoring cannot detect Attack~3. Finally, extending control-aware attack analysis to autopilots that use optimal or adaptive control approaches~\cite{lewis2012optimal,lewis2009reinforcement}, as well as to PX4, is a natural direction for future work.

\section{Conclusion}
\label{sec:conclusion}

This paper presented six control-aware attacks against ArduPilot that use legitimate, well-formed MAVLink commands to manipulate the internal dynamics of the autopilot. The attacks target different layers of ArduPilot's cascaded control architecture, including position guidance, PID rate loops, EKF state estimation, failsafe assumptions, yaw control, and combined multi-layer interactions. Evaluation in SITL and validation on a physical Pixhawk 2.4.8-based drone, as a subset of UAVs, show that short sequences of accepted MAVLink commands can produce measurable degradation in attitude stability, angular-rate behavior, trajectory tracking, and EKF health. When multiple weaknesses are combined, the drone can enter a self-sustaining unstable state that persists after the attacker stops transmitting and can lead to loss of control or crash.

The findings highlight a gap in current UAV defenses: Protocol-level protection alone is insufficient when trusted runtime interfaces can be used to issue unsafe control actions. Future defenses should therefore combine communication security with control-aware safeguards, including parameter range checks, limits on rapid parameter changes, EKF health monitoring, and detection logic that reasons about the physical and control-theoretic effects of command sequences. More broadly, this work shows the need to treat autopilot parameters and guidance commands as safety-critical inputs, not merely configuration data, and motivates extending control-aware attack analysis to other autopilots, drone types, and control architectures.

\vspace{.1in}
\noindent{\bf Disclaimer:} Ardupilot was informed about this paper. ArduPilot includes now a parameters lockdown mechanism, which when activated, mitigates the attacks described in the paper. 

\bibliographystyle{IEEEtran}
\bibliography{references}

\end{document}